\documentclass[conference, 10pt, letterpaper, oneside, draftcls, onecolumn]{IEEEtran} 
\usepackage{ifpdf}

\usepackage[noadjust]{cite}

\ifCLASSINFOpdf
  \usepackage[pdftex]{graphicx}
\else
  \usepackage[dvipdfmx]{graphicx}
\fi
\usepackage{amsmath}
\usepackage{array}

\ifCLASSOPTIONcompsoc
  \usepackage[caption=false, font=normalsize, labelfont=sf, textfont=sf]{subfig}
\else
  \usepackage[caption=false, font=footnotesize]{subfig}
\fi
\usepackage{fixltx2e}
\usepackage{url}

\usepackage{tikz}
\usepackage{tikz-3dplot}
\usetikzlibrary{positioning}
\usetikzlibrary{decorations.pathreplacing}
\usetikzlibrary{decorations.markings}
\usetikzlibrary{positioning}
\usetikzlibrary{shapes,arrows}
\tikzstyle{vecArrow} = [thick, decoration={markings,mark=at position
   1 with {\arrow[semithick]{open triangle 90}}},
   double distance=3pt, shorten >= 4pt,
   preaction = {decorate},
   postaction = {draw,line width=1.4pt, white,shorten >= 4.5pt}]
\tikzstyle{innerWhite} = [semithick, white,line width=1.4pt, shorten >= 4.5pt]

\usepackage{xcolor}
\definecolor{Green}{rgb}{0,0.5,0}
\definecolor{burgundy}{rgb}{0.545098,0,0}
\definecolor{burgundy}{rgb}{0.545098,0,0}
\definecolor{navyblue}{rgb}{0.0, 0.0, 0.5}
\definecolor{leafgreen}{rgb}{0.290196, 0.470588, 0.0}
\definecolor{bluegreen}{rgb}{0, 0.470588, 0.415686}
\definecolor{zuhl}{rgb}{0.1875, 0.26171875, 0.46484375}
\definecolor{orange}{rgb}{1, 0.6470588235, 0}

\usepackage{amsthm}
\usepackage{mathtools}
\usepackage{amssymb}
\usepackage{bm, bbm}
\usepackage{array}
\usepackage{xfrac}

\theoremstyle{plain}
\newtheorem{definition}{Definition}
\newtheorem{lemma}{Lemma}

\newtheorem{theorem}{Theorem}
\newtheorem{corollary}{Corollary}
\newtheorem{remark}{Remark}

\newtheorem{example}{Example}
\newtheorem{proposition}{Proposition}

\newcommand{\bvec}[1]{\boldsymbol{#1}}

\newcommand{\sgn}{\operatorname{sgn}}

\newcommand{\1}{\mbox{1}\hspace{-0.25em}\mbox{l}}
\newcommand{\lcm}{\mathrm{lcm}}

\newcommand{\figref}[1]{Fig.~\ref{#1}}
\newcommand{\lemref}[1]{Lemma~\ref{#1}}
\newcommand{\thref}[1]{Theorem~\ref{#1}}
\newcommand{\defref}[1]{Definition~\ref{#1}}
\newcommand{\corref}[1]{Corollary~\ref{#1}}
\newcommand{\sectref}[1]{Section~\ref{#1}}

\newcommand{\propref}[1]{Proposition~\ref{#1}}

\allowdisplaybreaks[3]

\usepackage[utf8]{inputenc}
\usepackage[font = footnotesize]{caption}
\usepackage{overpic}
\usepackage{pict2e}
\usepackage{lipsum}
\usepackage{docmute}
\usepackage{flushend}

\ifCLASSINFOpdf
	\usepackage[colorlinks]{hyperref} 
	\usepackage{microtype} 
\else
	\usepackage[colorlinks, dvipdfmx]{hyperref}
\fi

\begin{document}
%
\title{A Generalized Erasure Channel in the Sense of Polarization for Binary Erasure Channels}

\author{
\IEEEauthorblockN{Yuta Sakai and Ken-ichi Iwata}
\IEEEauthorblockA{Graduate School of Engineering, University of Fukui,\\
3-9-1 Bunkyo, Fukui, Fukui, 910-8507, Japan,\\
E-mail: \{y-sakai,~k-iwata\}@u-fukui.ac.jp}
}


%


\maketitle

\begin{abstract}
The polar transformation of a binary erasure channel (BEC) can be exactly approximated by other BECs.
Ar{\i}kan proposed that polar codes for a BEC can be efficiently constructed by using its useful property.
This study proposes a new class of arbitrary input generalized erasure channels, which can be exactly approximated the polar transformation by other same channel models, as with the BEC.
One of the main results is the recursive formulas of the polar transformation of the proposed channel.
In the study, we evaluate the polar transformation by using the $\alpha$-mutual information.
Particularly, when the input alphabet size is a prime power, we examines the following: (i) inequalities for the average of the $\alpha$-mutual information of the proposed channel after the one-step polar transformation, and (ii) the exact proportion of polarizations of the $\alpha$-mutual information of proposed channels in infinite number of polar transformations.
\end{abstract}


%
\IEEEpeerreviewmaketitle

\section{Introduction}

Polar codes were proposed by Ar{\i}kan \cite{arikan} as capacity achieving codes with low coding complexities for binary-input symmetric discrete memoryless channels (DMCs).
Previously, polar codes were generalized from binary codes to $q$-ary codes, $q \ge 3$.
The studies of $q$-ary polar codes are broadly divided into the following two approaches:
The first approach is the \emph{strong} polarization \cite{sasoglu, sasoglu2, mori}, i.e., the DMC is polarized to either \emph{noiseless} or \emph{pure noisy} channel. 
The second approach is the \emph{weak} polarization \cite{park2, sahebi, guo, nasser1, nasser2, nasser3}, i.e., the DMC is polarized to \emph{partially noiseless} channels.
The weak polarization is also called the \emph{multilevel polarization} \cite{nasser1, park2, sahebi}.
In this study, we consider the weak polarization.

To construct polar codes, channel parameters of virtual channels, generated by the polar transformation, are needed.
Commonly used channel parameters are the symmetric capacity and the Bhattacharyya parameter;
however, these computational complexities grow double-exponentially with the number of polar transformations.
In the binary-input case, Tal and Vardy \cite{tal} proposed an efficient solution to these computational complexities by approximating the polar transformation for each time.
In particular, it is known that the polar transformation of the binary erasure channel (BEC) can be exactly approximated by other BECs (cf. \propref{prop:BEC}).
Similarly, it is also known that the polar transformation of the $q$-ary ordered erasure channel ($q$-OEC), defined by Park and Barg \cite[p.~2285]{park1}, can be also exactly approximated by other $q$-OECs when $q$ is a power of two (cf. \cite[Section~III]{park2}).
Furthermore, Sahebi and Pradhan showed recursive formulas \cite[Eqs.~(3) and~(4)]{sahebi} of the polar transformation of the senary-input channel, defined in \cite[Fig.~4: Channel~2]{sahebi}.

In this study, we propose a new class of $q$-ary input DMCs $V$, as \defref{def:V} in \sectref{sect:V}. 
Since \defref{def:V} contains BECs, $q$-OECs, and \cite[Fig.~4: Channel~2]{sahebi}, the proposed channel $V$ is defined as a generalization of these channels.
To evaluate the polar transformation, we employ the $\alpha$-mutual information \cite{verdu} with the uniform input distribution, as defined in \eqref{def:symmetric_alpha}, rather than the symmetric capacity, the Bhattacharyya parameter, and Gallager's $E_{0}$ function \cite{red}. 
In the paper, we call it the symmetric capacity of order $\alpha$.
One of the main results is shown in \thref{th:V}, which gives the recursive formulas of the polar transformation of the proposed channel $V$, as with \propref{prop:BEC}.
Moreover, when the input alphabet size $q$ is a prime power, we investigate the polar transformation of the proposed channel $V$ in more detail.
Then, we derive the following two results: (i) inequalities for the average of the symmetric capacity of order $\alpha$ after the one-step polar transformation (cf. \corref{cor:ineq} and \figref{fig:alpha}), and
(ii) the exact proportion of the convergences of the symmetric capacity of order $\alpha$ in infinite number of polar transformations (cf. \thref{th:martingale} and \figref{fig:V27}).

\section{Preliminaries}

\subsection{Discrete memoryless channels and channel parameters}

Consider the DMC as follows:
For an integer $q \ge 2$, let $\mathbb{Z}_{q} \coloneqq \{ 0, 1, \dots, q-1 \}$ and $\mathcal{Y}$ be the input and output alphabets, respectively, where $\mathbb{Z}_{q}$ is called a complete residue system modulo $q$.
Note that the input alphabet size is denoted by $q$.
Then, the DMC $W : \mathbb{Z}_{q} \to \mathcal{Y}$ consists of a transition probability distribution $\{ W(y \mid x) \mid (x, y) \in \mathbb{Z}_{q} \times \mathcal{Y} \}$.
In this study, the input distribution $P_{X}$ is restricted to the uniform distribution on $\mathbb{Z}_{q}$, i.e., $P_{X}(x) = 1/q$ for all $x \in \mathbb{Z}_{q}$. 

We first introduce four kinds of channel parameters for DMCs.
Let $\ln$ denote the natural logarithm.
The symmetric capacity of $W : \mathbb{Z}_{q} \to \mathcal{Y}$ is denoted by
\begin{align}
I(W)
\coloneqq
\sum_{y \in \mathcal{Y}} \sum_{x \in \mathbb{Z}_{q}} \frac{1}{q} W(y \mid x) \ln \frac{ W(y \mid x) }{ \sum_{x^{\prime} \in \mathbb{Z}_{q}} (1/q) W(y \mid x^{\prime}) } ,
\label{def:symmetricI}
\end{align}
which is the mutual information between the input and output of $W$ under the uniform input distribution. 
In the channel coding theorem \cite{shannon}, it is shown that $I(W)$ is the supremum of achievable rates through the DMC $W$ under the uniform input distribution.
In uncoded schemes, the average probability of error with a maximum likelihood decoder is calculated by
\begin{align}
P_{\mathrm{e}}( W )
\coloneqq
1 - \sum_{y \in \mathcal{Y}} \frac{1}{q} \max_{x \in \mathbb{Z}_{q}} W(y \mid x)
\end{align}
for $W : \mathbb{Z}_{q} \to \mathcal{Y}$, where note that $0 \le P_{\mathrm{e}}(W) \le (q-1)/q$ for any $W$.
Moreover, the average Bhattacharyya distance of $W$, defined by {\c{S}}a{\c{s}}o{\u{g}}lu et al. \cite[Eq.~(7)]{sasoglu}, is denoted by
\begin{align}
Z(W)
& \coloneqq
\frac{ 1 }{ q (q-1) } \sum_{\substack{ x, x^{\prime} \in \mathbb{Z}_{q} : \\ x \neq x^{\prime} }} \sum_{y \in \mathcal{Y}} \sqrt{ W(y \mid x) W(y \mid x^{\prime}) } ,
\end{align}
which is used to bounds on $P_{\mathrm{e}}( W )$ and $I( W )$ (cf. \cite[Propositions~2 and~3]{sasoglu}).
Furthermore, the $E_{0}$ function of $W$, defined by Gallager \cite[Eq.~(5.6.14)]{red}, with the uniform input distribution is denoted by
\begin{align}
E_{0}(\rho, W)
\coloneqq
- \ln \Bigg[ \sum_{y \in \mathcal{Y}} \Bigg( \sum_{x \in \mathbb{Z}_{q}} \frac{1}{q} W(y \mid x)^{1/(1+\rho)} \Bigg)^{1+\rho} \Bigg]
\label{def:E0}
\end{align}
for $\rho \in (-1, \infty)$, which is used in error exponents for DMCs (cf. \cite{red, sphere, arimoto}).
Note that Alsan and Telatar \cite{alsan2} investigated that Ar{\i}kan's original polar transformation \cite[Eqs.~(17) and~(18)]{arikan} for binary-input DMCs $W$ increases the average of the $E_{0}$ functions for each $\rho \ge 0$.

Instead of the channel parameters \eqref{def:symmetricI}--\eqref{def:E0}, in this study, we use the symmetric capacity of order $\alpha$, which is defined by
\begin{align}
I_{\alpha}( W )
\coloneqq
\frac{ \alpha }{ \alpha-1 } \ln \Bigg[ \sum_{y \in \mathcal{Y}} \Bigg( \sum_{x \in \mathbb{Z}_{q}} \frac{1}{q} W(y \mid x)^{\alpha} \Bigg)^{1/\alpha} \Bigg]
\label{def:symmetric_alpha}
\end{align}
for a channel $W : \mathbb{Z}_{q} \to \mathcal{Y}$ and $\alpha \in (0, 1) \cup (1, \infty)$, where the quantity \eqref{def:symmetric_alpha} is identical to the mutual information of order $\alpha$, defined by Arimoto \cite[Eq.~(16)]{arimoto2}, under the uniform input distribution.
In addition, the quantity \eqref{def:symmetric_alpha} is also identical to the $\alpha$-mutual information \cite[Eq.~(53)]{verdu} under uniform input distribution, and it was recently studied by Ho and Verd{\'u} \cite{ho}.
Following \cite[Theorem~4]{ho}, for $\alpha \in \{ 0, 1, \infty \}$, we also define the symmetric capacity of order $\alpha$ as follows:
\begin{align}
I_{0}( W )
& \coloneqq \!
\lim_{\alpha \to 0^{+}} \! I_{\alpha}( W )
=
\min_{y \in \mathcal{Y}} \bigg( \! \ln \frac{q}{ |\{ x \in \mathbb{Z}_{q} \mid W(y \mid x) > 0 \} | } \bigg) ,
\label{def:alpha_0} \\
I_{1}( W )
& \coloneqq \!
\lim_{\alpha \to 1} I_{\alpha}( W )
=
I( W ) ,
\label{def:alpha_1} \\
I_{\infty}( W )
& \coloneqq \!
\lim_{\alpha \to \infty} I_{\alpha}( W )
=
\ln \Bigg( \sum_{y \in \mathcal{Y}} \max_{x \in \mathbb{Z}_{q}} W(y \mid x) \Bigg) ,
\label{def:alpha_infty}
\end{align}
where $| \cdot |$ denotes the cardinality of the finite set.
We now readily see the following identities:
\begin{align}
I_{\alpha}(W)
& =
\frac{\alpha}{1-\alpha} E_{0} \bigg( \frac{1 - \alpha}{\alpha}, W \bigg)
\qquad \mathrm{for} \ \alpha \in (0, 1) \cup (1, \infty) ,
\label{eq:I_E0} \\
I_{1/2}(W)
& =
E_{0}(1, W)
=
\ln \frac{q}{1 + (q-1) Z(W)} ,
\\
I_{\infty}(W)
& =
(\ln q) + \ln \Big( 1 - P_{\mathrm{e}}(W) \Big) ,
\end{align}
where $E_{0}(1, W)$ is called the (symmetric) cutoff rate.
Thus, the symmetric capacity of order $\alpha$, denoted by $I_{\alpha}(W)$, is closely related to  $I(W)$, $P_{\mathrm{e}}( W )$, $Z( W )$, and $E_{0}(\rho, W)$;
and therefore, we employ $I_{\alpha}(W)$ for $\alpha \in [0, \infty]$ to evaluate the channel parameters \eqref{def:symmetricI}--\eqref{def:E0} in the study.

\subsection{Polar transformations for $q$-ary input channels with $q \ge 2$}

For $\gamma \in \mathbb{Z}_{q}$, we define the mapping $f_{\gamma} : \mathbb{Z}_{q}^{2} \to \mathbb{Z}_{q}$ as
\begin{align}
f_{\gamma}(u_{1}, u_{2})
\coloneqq
u_{1} \oplus (\gamma \otimes u_{2}) ,
\label{def:f}
\end{align}
where $\oplus$ and $\otimes$ denote the addition and multiplication modulo $q$, respectively.
Using the mapping $f_{\gamma} : \mathbb{Z}_{q}^{2} \to \mathbb{Z}_{q}$, we now introduce the one-step polar transformation used in the study as follows.
For a DMC $W : \mathbb{Z}_{q} \to \mathcal{Y}$, the one-step polar transformation creates the channel $W^{0} : \mathbb{Z}_{q} \to \mathcal{Y}^{2}$ as
\begin{align}
W^{0}(y_{1}, y_{2} \mid u_{1})
& \coloneqq
\sum_{u_{2}^{\prime} \in \mathbb{Z}_{q}} \frac{1}{q} W(y_{1} \mid f_{\gamma}(u_{1}, u_{2}^{\prime})) W(y_{2} \mid u_{2}^{\prime}) ,
\label{def:W0}
\end{align}
and the channel $W^{1} : \mathbb{Z}_{q} \to \mathcal{Y}^{2} \times \mathbb{Z}_{q}$ as
\begin{align}
W^{1}(y_{1}, y_{2}, u_{1} \mid u_{2})
& \coloneqq
\frac{1}{q} W(y_{1} \mid f_{\gamma}(u_{1}, u_{2})) W(y_{2} \mid u_{2}) .
\label{def:W1}
\end{align}
In the paper, the above polar transformation is denoted by $W \overset{\gamma}{\mapsto} (W^{0}, W^{1})$.
Note that, if $\gamma = 1$, then the polar transformation $W \overset{1}{\mapsto} (W^{0}, W^{1})$ is reduced to the polar transformation discussed by Park and Barg \cite{park2} and Sahebi and Pradhan \cite{sahebi}.

\begin{remark}
We now consider the binary operation $\ast$ on $\mathbb{Z}_{q}$ as $u_{1} \ast u_{2} \coloneqq f_{\gamma}(u_{1}, u_{2})$, defined in \eqref{def:f}. 
Let $\gcd(a, b)$ denote the greatest common divisor of $a, b \in \mathbb{N}$, and let $\mathbb{Z}_{q}^{\times} \coloneqq \{ z \in \mathbb{Z}_{q} \mid \gcd( z, q ) = 1 \}$ denote a reduced residue system modulo $q$.
If $\gamma \in \mathbb{Z}_{q}^{\times}$, then it is easy to see that $(\mathbb{Z}_{q}, \ast)$ forms a quasigroup (cf. \cite[Definition~1]{nasser3}).
Therefore, it follows from \cite[Theorem~1]{nasser3} that the polar transformation $W \overset{\gamma}{\mapsto} (W^{0}, W^{1})$ with $\gamma \in \mathbb{Z}_{q}^{\times}$ behaves the weak polarization.
However, if $\gamma \notin \mathbb{Z}_{q}^{\times}$, then $(\mathbb{Z}_{q}, \ast)$ does not form a quasigroup in general.
Thus, in this paper, we only consider the mapping $f_{\gamma} : \mathbb{Z}_{q}^{2} \to \mathbb{Z}_{q}$ for $\gamma \in \mathbb{Z}_{q}^{\times}$.
\end{remark}

Let $\mathbb{N} \coloneqq \{ 1, 2, \dots \}$ be the set of positive integers, and let $\mathbb{N}_{0} \coloneqq \mathbb{N} \cup \{ 0 \}$ be the set of nonnegative integers.
After the $n$-step polar transformations of a channel $W$ for $n \in \mathbb{N}$, the channel $W_{2^{n}}^{(i)} : \mathbb{Z}_{q} \to \mathcal{Y}^{2^{n}} \times \mathbb{Z}_{q}^{i}$ is created by
\begin{align}
W_{2^{n}}^{(i)}
\coloneqq
\big( \cdots \big( \big( W^{b_{1}} \big)^{b_{2}} \big)^{b_{3}} \cdots \big)^{b_{n}}
\label{def:Wi}
\end{align}
for each $i \in \{ 0, 1, \dots, 2^{n}-1 \}$, where suppose that $W_{0}^{(0)} \coloneqq W$, and $(b_{1}, b_{2}, \dots, b_{n}) \in \{ 0, 1 \}^{n}$ is the binary representation of the index $i$.
Note that the most significant bit of $(b_{1}, b_{2}, \dots, b_{n})$ is $b_{1}$.
The channel $W_{2^{n}}^{(i)}$ is sometimes called the virtual channel.
Since the output alphabet size $\big| \mathcal{Y}^{2^{n}} \times \mathbb{Z}_{q}^{i} \big|$ of the channel $W_{2^{n}}^{(i)}$ grows double-exponentially with the number $n$ of polar transformations, the computation of $I_{\alpha} \big( W_{2^{n}}^{(i)} \big)$ turns out to be complicate with increasing $n$.
This computational difficulty is a main factor that polar codes are hard to construct in general.

Fortunately, the quantity $I_{\alpha}\big( W_{2^{n}}^{(i)} \big)$ can be easily calculated when $W$ is a BEC.
To see this computational simplicity, we now define an equivalence relation of $q$-ary input DMCs in the sense of the polar transformations as follows:

\begin{definition}
\label{def:equiv}
A $q$-ary input channel $W$ is said to be equivalent to another $q$-ary input channel $W^{\prime}$ if
\begin{align}
I_{\alpha} \Big( W_{2^{n}}^{(i)} \Big)
& =
I_{\alpha} \Big( (W^{\prime})_{2^{n}}^{(i)} \Big)
\end{align}
for any $n \in \mathbb{N}_{0}$, $i \in \{ 0, 1, \dots, 2^{n} - 1 \}$, and $\alpha \in [0, \infty]$.
\end{definition}

\defref{def:equiv} means that, if $W$ is equivalent to $W^{\prime}$, then both of these polar transformations are identically behaved in the sense of the symmetric capacity of order $\alpha$.
Note that the equivalence of \defref{def:equiv} is given in a similar sense to \cite[Eqs.~(2)~and~(3)]{mori}.
We now consider the BEC $W_{\mathrm{BEC}} : \{ 0, 1 \} \to \{ 0, 1, ? \}$ with an erasure probability $\varepsilon \in [0, 1]$ as
\begin{align}
W_{\mathrm{BEC}}(y \mid x)
\coloneqq
\begin{cases}
1 - \varepsilon
& \mathrm{if} \ y = x , \\
\varepsilon
& \mathrm{if} \ y = \; ? , \\
0
& \mathrm{otherwise} .
\end{cases}
\label{def:bec}
\end{align}
For simplicity, we denote by $\mathrm{BEC}( \varepsilon )$ the BEC which the erasure probability is $\varepsilon \in [0, 1]$.
It is known that the polar transformation of BECs can be easily analyzed, as shown in the following proposition.

\begin{proposition}[{\cite[Proposition~6]{arikan}}]
\label{prop:BEC}
Consider the polar transformation $W_{\mathrm{BEC}} \overset{1}{\mapsto} (W_{\mathrm{BEC}}^{0}, W_{\mathrm{BEC}}^{1})$ for $W_{\mathrm{BEC}} \equiv \mathrm{BEC}( \varepsilon )$.
Then, the channels $W_{\mathrm{BEC}}^{0}$ and $W_{\mathrm{BEC}}^{1}$ are equivalent to $\mathrm{BEC}( 2 \varepsilon - \varepsilon^{2} )$ and $\mathrm{BEC}( \varepsilon^{2} )$, respectively.
\end{proposition}

\propref{prop:BEC} shows that the polar transformation of a BEC can be exactly approximated by other BECs again.
Therefore, it is sufficient to propagate the erasure probability $\varepsilon$ recursively with the formulas $2 \varepsilon - \varepsilon^{2}$ and $\varepsilon^{2}$.

\section{A class of generalized erasure channels $V$}
\label{sect:V}

In this section, we propose a generalization of erasure channels, and consider the polar transformations of its channel.
Let $\mathbb{Z} \coloneqq \{ \dots, -2, -1, 0, 1, 2, \dots \}$ be the set of integers.
The set of integers modulo $m \in \mathbb{N}$ is denoted by $\mathbb{Z} / m \mathbb{Z} \coloneqq \{ [ r ]_{m} \mid r \in \mathbb{Z} \}$, where $[ r ]_{m} \coloneqq \{ z \in \mathbb{Z} \mid z \equiv r \pmod{m} \}$ is the residue class of $r \in \mathbb{Z}$ modulo $m$.
Moreover, let $\mathcal{D}_{m} \coloneqq \{ d \in \mathbb{N} \mid d \equiv 0 \pmod{m} \}$ be the set of positive divisors of $m \in \mathbb{N}$.
When we denote by $( x_{i} : i \in \mathcal{I} )$ a vector with an index set $\mathcal{I}$, we define the proposed channel $V$ of the study as follows:

\begin{definition}
\label{def:V}
If the input alphabet size is $q$, then the channel $V : \mathbb{Z}_{q} \to \mathcal{Y}$ is defined by $\mathcal{Y} = \bigcup_{d \in \mathcal{D}_{q}} \big( \mathbb{Z} / d \mathbb{Z} \big)$ and
\begin{align}
V(y \mid x)
\coloneqq
\begin{cases}
\varepsilon_{d}
& \mathrm{if} \ y = [x]_{d} \ \mathrm{for} \ \mathrm{some} \ d \in \mathcal{D}_{q} , \\
0
& \mathrm{otherwise}
\end{cases}
\label{def:eq:V}
\end{align}
for a given probability vector $( \varepsilon_{d} : d \in \mathcal{D}_{q} )$, where note that $\sum_{d \in \mathcal{D}_{q}} \varepsilon_{d} = 1$ and $\varepsilon_{d} \ge 0$ for all $d \in \mathcal{D}_{q}$.
\end{definition}

Note that the channel $V$ of \defref{def:V} is symmetric (cf. \cite[p.~94]{red}).
As with the notation $\mathrm{BEC}( \varepsilon )$, we also denote by $V_{(q)}( \varepsilon_{d} : d \in \mathcal{D}_{q} )$ the $q$-ary input channel $V$ with the probability vector $( \varepsilon_{d} : d \in \mathcal{D}_{q} )$.
It is easy to see that, if the input alphabet is binary, i.e., $q = 2$, then $V_{(2)}( \varepsilon_{1}, \varepsilon_{2} )$ is reduced to $\mathrm{BEC}( \varepsilon_{1} )$.
Hence, it follows that \defref{def:V} contains the BEC{\@}.
In addition, if the input alphabet size $q$ is a prime number, then $V_{(q)}(\varepsilon_{1}, \varepsilon_{q})$ is also reduced to the (conventional) $q$-ary erasure channel (cf. \cite[p.~589]{mackay}). 
The following example shows the senary-input channel $V$ of \defref{def:V}, i.e., when the input alphabet size $q = 6 = 2 \cdot 3$ is composite.

\begin{example}
In the case of $q = 6$, we now check an example of \defref{def:V}.
The input alphabet is $\mathbb{Z}_{6} = \{ 0, 1, 2, 3, 4, 5 \}$.
Since $\mathcal{D}_{6} = \{ 1, 2, 3, 6 \}$, the output alphabet is $\mathcal{Y} = \big( \mathbb{Z} / 1 \mathbb{Z} \big) \cup \big( \mathbb{Z} / 2 \mathbb{Z} \big) \cup \big( \mathbb{Z} / 3 \mathbb{Z} \big) \cup \big( \mathbb{Z} / 6 \mathbb{Z} \big) = \{ [0]_{1} \} \cup \{ [0]_{2}, [1]_{2} \} \cup \{ [0]_{3}, [1]_{3}, [2]_{3} \} \cup \{ [0]_{6}, [1]_{6}, [2]_{6}, [3]_{6}, [4]_{6}, [5]_{6} \}$.
Therefore, the channel $V : \mathbb{Z}_{6} \to \mathcal{Y}$ is given by
\begin{align}
V(y \mid x)
=
\begin{cases}
\varepsilon_{6}
& \mathrm{if} \ y = [x]_{6} , \\
\varepsilon_{3}
& \mathrm{if} \ y = [x]_{3} , \\
\varepsilon_{2}
& \mathrm{if} \ y = [x]_{2} , \\
\varepsilon_{1}
& \mathrm{if} \ y = [x]_{1} \ (= \mathbb{Z}) , \\
0
& \mathrm{otherwise}
\end{cases}
\label{eq:6ary_V}
\end{align}
for a 4-dimensional probability vector $( \varepsilon_{d} : d \in \mathcal{D}_{6} ) = (\varepsilon_{1}, \varepsilon_{2}, \varepsilon_{3}, \varepsilon_{6})$.
This channel is denoted by $V_{(6)}( \varepsilon_{1}, \varepsilon_{2}, \varepsilon_{3}, \varepsilon_{6} )$, and it is identical to \cite[Fig.~4: Channel~2]{sahebi}.
\end{example}

Furthermore, it can be seen that \defref{def:V} contains the $q$-OEC, defined by Park and Barg \cite[p.~2285]{park1}, when the input alphabet size $q$ is a prime power, i.e., $q = p^{m}$ for some prime $p$ and some $m \in \mathbb{N}$.

We now consider the polar transformation for the channel $V$.
Let $\lcm(a, b)$ be the least common multiple of $a, b \in \mathbb{N}$.
The following theorem shows that the channel $V$ has an useful property in terms of the polar transformation, as with the BEC (cf. \propref{prop:BEC}). %

\begin{theorem}
\label{th:V}
Consider the polar transformation $V \overset{\gamma}{\mapsto} (V^{0}, V^{1})$ for $V \equiv V_{(q)}( \varepsilon_{d} : d \in \mathcal{D}_{q} )$.
If $\gamma \in \mathbb{Z}_{q}^{\times} \coloneqq \{ z \in \mathbb{Z}_{q} \mid \gcd(z, q) = 1 \}$, then the channels $V^{0}$ and $V^{1}$ are equivalent to $V_{(q)}( \varepsilon_{d}^{-} : d \in \mathcal{D}_{q} )$ and $V_{(q)}( \varepsilon_{d}^{+} : d \in \mathcal{D}_{q} )$, respectively, where $( \varepsilon_{d}^{-} : d \in \mathcal{D}_{q} )$ and $( \varepsilon_{d}^{+} : d \in \mathcal{D}_{q} )$ are given by 
\begin{align}
\varepsilon_{d}^{-}
=
\sum_{\substack{ d_{1}, d_{2} \in \mathcal{D}_{q} : \\ \gcd( d_{1}, d_{2} ) = d }} \varepsilon_{d_{1}} \, \varepsilon_{d_{2}}
\qquad \mathrm{and} \qquad
\varepsilon_{d}^{+}
=
\sum_{\substack{ d_{1}, d_{2} \in \mathcal{D}_{q} : \\ \lcm( d_{1}, d_{2} ) = d }} \varepsilon_{d_{1}} \, \varepsilon_{d_{2}}
\label{eq:E+-}
\end{align}
for $d \in \mathcal{D}_{q}$, respectively.
\end{theorem}

\begin{IEEEproof}[Proof of \thref{th:V}]
We first introduce a part of the Chinese reminder theorem as follows.

\begin{lemma}[Chinese reminder theorem]
\label{lem:chinese}
For any $z_{1}, z_{2} \in \mathbb{Z}$ and $d_{1}, d_{2} \in \mathbb{N}$, the system of two congruences
\begin{align}
z_{1} & \equiv r^{\prime}
\pmod{d_{1}} ,
\\
z_{2} & \equiv r^{\prime}
\pmod{d_{2}}
\end{align}
has a unique solution $r^{\prime} \in \mathbb{Z}$ modulo $\lcm(d_{1}, d_{2})$ if and only if $z_{1} \equiv z_{2} \pmod{\gcd( d_{1}, d_{2} )}$ holds.
\end{lemma}

In addition, we present the following lemma.

\begin{lemma}
\label{lem:bin}
For a channel $W : \mathbb{Z}_{q} \to \mathcal{Y}$, we define an output subalphabet
\begin{align}
\mathcal{B}( y )
\coloneqq
\big\{ y^{\prime} \in \mathcal{Y} \mid \forall x \in \mathbb{Z}_{q}, P_{X|Y}(x \mid y^{\prime}) = P_{X|Y}(x \mid y) \big\}
\label{def:bin}
\end{align}
for $y \in \mathcal{Y}$, where $P_{X|Y}$ is the a posteriori probability distribution of $W$.
Moreover, the channel $W_{\ast} : \mathbb{Z}_{q} \to \mathcal{Y}_{\ast}$ with respect to $W$ is defined by
\begin{align}
W_{\ast}(\mathcal{B}( y ) \mid x)
\coloneqq
\sum_{y^{\prime} \in \mathcal{B}( y )} W(y^{\prime} \mid x)
\label{def:W_ast}
\end{align}
for $(x, \mathcal{B}( y )) \in \mathbb{Z}_{q} \times \mathcal{Y}_{\ast}$, where the output alphabet $\mathcal{Y}_{\ast}$ is given by
$
\mathcal{Y}_{\ast}
\coloneqq
\{ \mathcal{B}( y ) \mid y \in \mathcal{Y} \}
$.
Then, the channel $W$ is equivalent to the channel $W_{\ast}$ in the sense of \defref{def:equiv}.
\end{lemma}

\lemref{lem:bin} can be derived from the equivalence relation $\overset{\mathrm{i}}{\sim}$, discussed in \cite[p.~2722]{mori}.
\lemref{lem:bin} implies that groups of output symbols, which have same a posteriori distribution, can be merged into one symbol.

Let $q \ge 2$ be a fixed integer.
For a pair of uniformly distributed random variables $(U_{1}, U_{2}) \in \mathbb{Z}_{q}^{2}$, i.e., the pair of random variables $(U_{1}, U_{2})$ follows the uniform distribution
\begin{align}
P_{U_{1}, U_{2}}( u_{1}, u_{2} )
=
\frac{1}{q^{2}}
\label{eq:unif_u}
\end{align}
for $(u_{1}, u_{2}) \in \mathbb{Z}_{q}^{2}$, we define the pair of random variables $(X_{1}, X_{2})$ as
\begin{align}
(X_{1}, X_{2})
\coloneqq
(f_{\gamma}(U_{1}, U_{2}), U_{2}) ,
\label{def:x1x2}
\end{align}
where the mapping $f_{\gamma} : \mathbb{Z}_{q}^{2} \to \mathbb{Z}_{q}$ is defined in \eqref{def:f} for a fixed $\gamma \in \mathbb{Z}_{q}$.
That is, we consider the following system:
\begin{align}
U_{1} + \gamma \, U_{2} & \equiv X_{1} \pmod{q} ,
\label{eq:system_ux_1} \\
U_{2} & \equiv X_{2} \pmod{q} .
\label{eq:system_ux_2}
\end{align}
In \eqref{def:x1x2}, suppose throughout the proof that $\gamma \in \mathbb{Z}_{q}^{\times} \coloneqq \{ z \in \mathbb{Z}_{q} \mid \gcd(z, q) = 1 \}$.
Since the congruence $\gamma \, u \equiv x \pmod{q}$ has a unique solution $u$ modulo $q$ for a given $x$ if $\gcd( \gamma, q ) = 1$, it follows from \eqref{eq:system_ux_1} and \eqref{eq:system_ux_2} that the pairs $(U_{1}, U_{2}) \in \mathbb{Z}_{q}^{2}$ and $(X_{1}, X_{2}) \in \mathbb{Z}_{q}^{2}$ are in one to one correspondence when $\gamma \in \mathbb{Z}_{q}^{\times}$. 
Moreover, since $(U_{1}, U_{2})$ follows the uniform distribution on $\mathbb{Z}_{q}^{2}$, the pair $(X_{1}, X_{2})$ also follows the uniform distribution on $\mathbb{Z}_{q}^{2}$ when $\gamma \in \mathbb{Z}_{q}^{\times}$.
Namely, both of the probability distributions $P_{X_{1}}$ of $X_{1}$ and $P_{X_{2}}$ of $X_{2}$ satisfy
\begin{align}
P_{X_{1}}( x_{1} ) = P_{X_{2}}( x_{2} ) = \frac{1}{q}
\label{eq:unif}
\end{align}
for $x_{1}, x_{2} \in \mathbb{Z}_{q}$.

Let $\mathcal{Y} = \bigcup_{d \in \mathcal{D}_{q}} \big( \mathbb{Z} / d \mathbb{Z} \big) = \big\{ [ r ]_{d} \mid d \in \mathcal{D}_{q} \ \mathrm{and} \ r \in \mathbb{Z}_{d} \big\}$.
For two random variables $X_{1}$ and $X_{2}$ generated by \eqref{def:x1x2}, we now consider two pairs of random variables $(X_{1}, Y_{1}), (X_{2}, Y_{2}) \in \mathbb{Z}_{q} \times \mathcal{Y}$ which follow the joint probability distribution
\begin{align}
P_{X_{1}, X_{2}, Y_{1}, Y_{2}}(x_{1}, x_{2}, y_{1}, y_{2})
& =
P_{X_{1}, Y_{1}}(x_{1}, y_{1}) P_{X_{2}, Y_{2}}(x_{2}, y_{2})
\label{eq:ind}
\end{align}
for $(x_{1}, x_{2}, y_{1}, y_{2}) \in \mathbb{Z}_{q}^{2} \times \mathcal{Y}^{2}$, where the conditional distributions $P_{Y_{1}|X_{1}}$ and $P_{Y_{2}|X_{2}}$ are given by
\begin{align}
P_{Y_{1}|X_{1}}(y_{1} \mid x_{1})
& \coloneqq
V(y_{1} \mid x_{1}) ,
\label{def:x1y1} \\
P_{Y_{2}|X_{2}}(y_{2} \mid x_{2})
& \coloneqq
V(y_{2} \mid x_{2})
\label{def:x2y2}
\end{align}
for $(x_{1}, y_{1}) \in \mathbb{Z}_{q} \times \mathcal{Y}$ and $(x_{2}, y_{2}) \in \mathbb{Z}_{q} \times \mathcal{Y}$, respectively, and the channel $V : \mathbb{Z}_{q} \to \mathcal{Y}$ is defined in \defref{def:V}.
Note that the independence between $(X_{1}, Y_{1})$ and $(X_{2}, Y_{2})$, as shown in \eqref{eq:ind}, comes from the following two hypotheses: (i) the channel $V$ is memoryless, and (ii) the channel $V$ is used without feedback.
Summing over all $(x_{1}, x_{2}) \in \mathbb{Z}_{q}^{2}$ for both sides of \eqref{eq:ind}, we see that
\begin{align}
P_{Y_{1}, Y_{2}}(y_{1}, y_{2})
=
P_{Y_{1}}(y_{1}) P_{Y_{2}}(y_{2})
\label{eq:outputs}
\end{align}
for $(y_{1}, y_{2}) \in \mathcal{Y}^{2}$.
Moreover, direct calculations show
\begin{align}
P_{Y_{1}}( [ r ]_{d} )
& =
\sum_{x \in \mathbb{Z}_{q}} P_{X_{1}, Y_{1}}(x, [ r ]_{d})
\\
& \overset{\eqref{eq:unif}}{=}
\sum_{x \in \mathbb{Z}_{q}} \frac{1}{q} P_{Y_{1}|X_{1}}([ r ]_{d} \mid x)
\\
& \overset{\eqref{def:x1y1}}{=}
\sum_{x \in \mathbb{Z}_{q}} \frac{ 1 }{ q } V( [ r ]_{d} \mid x )
\\
& \overset{\eqref{def:eq:V}}{=}
\sum_{x \in \mathbb{Z}_{q} \cap [ r ]_{d}} \frac{ \varepsilon_{d} }{ q }
\\
& \overset{\text{(a)}}{=}
\frac{ q }{ d } \times \frac{ \varepsilon_{d} }{ q }
\\
& =
\frac{ \varepsilon_{d} }{ d } ,
\label{eq:marginal_out} \\
P_{X_{1}|Y_{1}}(x \mid [r]_{d})
& =
\frac{ P_{X_{1}, Y_{1}}(x, [r]_{d}) }{ P_{Y_{1}}( [ r ]_{d} ) }
\\
& \overset{\eqref{eq:unif}}{=}
\bigg( \frac{ 1 }{ q } P_{X_{1}|Y_{1}}( [r]_{d} \mid x ) \bigg) \Bigg/ P_{Y_{1}}( [ r ]_{d} )
\\
& \overset{\eqref{def:x1y1}}{=}
\bigg( \frac{ 1 }{ q } V( [ r ]_{d} \mid x ) \bigg) \Bigg/ P_{Y_{1}}( [ r ]_{d} )
\\
& \overset{\eqref{eq:marginal_out}}{=}
\bigg( \frac{ 1 }{ q } V( [ r ]_{d} \mid x ) \bigg) \Bigg/ \bigg( \frac{ \varepsilon_{d} }{ d } \bigg)
\\
& =
\frac{ d }{ q } \times \frac{ V( [ r ]_{d} \mid x ) }{ \varepsilon_{d} }
\\
& \overset{\eqref{def:eq:V}}{=}
\begin{cases}
d/q 
& \mathrm{if} \ x \in [ r ]_{d} , \\
0
& \mathrm{otherwise}
\end{cases}
\label{eq:posteriori}
\end{align}
for $(x, [r]_{d}) \in \mathbb{Z}_{q} \times \mathcal{Y}$, where (a) follows by Lagrange's theorem.
We divide this proof into the following two parts:

\subsection{First part: Channel $V^{0}$}

In the first part, we consider the channel $V^{0} : \mathbb{Z}_{q} \to \mathcal{Y}^{2}$, which is generated by $V \overset{\gamma}{\mapsto} (V^{0}, V^{1})$, as
\begin{align}
V^{0}(y_{1}, y_{2} \mid u_{1})
& \overset{\eqref{def:W0}}{=}
\sum_{u_{2}^{\prime} \in \mathbb{Z}_{q}} \frac{1}{q} V(y_{1} \mid f_{\gamma}(u_{1}, u_{2}^{\prime})) V(y_{2} \mid u_{2}^{\prime})
\\
& \overset{\eqref{eq:unif}}{=}
\sum_{u_{2}^{\prime} \in \mathbb{Z}_{q}} P_{X_{2}}( u_{2}^{\prime} ) V(y_{1} \mid f_{\gamma}(u_{1}, u_{2}^{\prime})) V(y_{2} \mid u_{2}^{\prime})
\\
& \overset{\eqref{def:x1y1}}{=}
\sum_{u_{2}^{\prime} \in \mathbb{Z}_{q}} P_{X_{2}}( u_{2}^{\prime} ) P_{Y_{1}|X_{1}}(y_{1} \mid f_{\gamma}(u_{1}, u_{2}^{\prime})) V(y_{2} \mid u_{2}^{\prime})
\\
& \overset{\eqref{def:x2y2}}{=}
\sum_{u_{2}^{\prime} \in \mathbb{Z}_{q}} P_{X_{2}}( u_{2}^{\prime} ) P_{Y_{1}|X_{1}}(y_{1} \mid f_{\gamma}(u_{1}, u_{2}^{\prime})) P_{Y_{2}|X_{2}}(y_{2} \mid u_{2}^{\prime})
\\
& =
\sum_{u_{2}^{\prime} \in \mathbb{Z}_{q}} \frac{ P_{X_{1}}(f_{\gamma}(u_{1}, u_{2}^{\prime})) P_{X_{2}}( u_{2}^{\prime} ) P_{Y_{1}|X_{1}}(y_{1} \mid f_{\gamma}(u_{1}, u_{2}^{\prime})) P_{Y_{2}|X_{2}}(y_{2} \mid u_{2}^{\prime}) }{ P_{X_{1}}(f_{\gamma}(u_{1}, u_{2}^{\prime})) }
\\
& =
\sum_{u_{2}^{\prime} \in \mathbb{Z}_{q}} \frac{ P_{X_{1}, Y_{1}}(f_{\gamma}(u_{1}, u_{2}^{\prime}), y_{1}) P_{X_{2}, Y_{2}}(u_{2}^{\prime}, y_{2}) }{ P_{X_{1}}(f_{\gamma}(u_{1}, u_{2}^{\prime})) }
\\
& \overset{\eqref{eq:ind}}{=}
\sum_{u_{2}^{\prime} \in \mathbb{Z}_{q}} \frac{ P_{X_{1}, X_{2}, Y_{1}, Y_{2}}(f_{\gamma}(u_{1}, u_{2}^{\prime}), u_{2}^{\prime}, y_{1}, y_{2}) }{ P_{X_{1}}(f_{\gamma}(u_{1}, u_{2}^{\prime})) }
\\
& \overset{\eqref{def:x1x2}}{=}
\sum_{u_{2}^{\prime} \in \mathbb{Z}_{q}} \frac{ P_{U_{1}, U_{2}, Y_{1}, Y_{2}}(u_{1}, u_{2}^{\prime}, y_{1}, y_{2}) }{ P_{U_{1}}(u_{1}) }
\\
& =
\sum_{u_{2}^{\prime} \in \mathbb{Z}_{q}} P_{U_{2}, Y_{1}, Y_{2} \mid U_{1}}(u_{2}^{\prime}, y_{1}, y_{2} \mid u_{1})
\\
& =
P_{Y_{1}, Y_{2} | U_{1}}(y_{1}, y_{2} \mid u_{1})
\label{eq:V0}
\end{align}
for $(u_{1}, y_{1}, y_{2}) \in \mathbb{Z}_{q} \times \mathcal{Y}^{2}$.
From \eqref{eq:outputs} and \eqref{eq:marginal_out}, we get
\begin{align}
\min\{ \varepsilon_{d_{1}}, \varepsilon_{d_{2}} \} > 0
\iff
P_{Y_{1}, Y_{2}}( [r_{1}]_{d_{1}}, [r_{2}]_{d_{2}} ) > 0
\label{iff:outputs}
\end{align}
for $( [r_{1}]_{d_{1}}, [r_{2}]_{d_{2}} ) \in \mathcal{Y}^{2}$.
With attention to \eqref{iff:outputs}, the a posteriori probability of \eqref{eq:V0} is given by
\begin{align}
P_{U_{1}|Y_{1}, Y_{2}}(u_{1} \mid y_{1}, y_{2})
& =
\frac{ P_{U_{1}, Y_{1}, Y_{2}}(u_{1}, y_{1}, y_{2}) }{ P_{Y_{1}, Y_{2}}(y_{1}, y_{2}) }
\\
& =
\frac{ \sum_{u_{2}^{\prime} \in \mathbb{Z}_{q}} P_{U_{1}, U_{2}, Y_{1}, Y_{2}}(u_{1}, u_{2}^{\prime}, y_{1}, y_{2}) }{ P_{Y_{1}, Y_{2}}(y_{1}, y_{2}) }
\\
& \overset{\eqref{def:x1x2}}{=}
\frac{ \sum_{u_{2}^{\prime} \in \mathbb{Z}_{q}} P_{X_{1}, X_{2}, Y_{1}, Y_{2}}(f_{\gamma}( u_{1}, u_{2}^{\prime} ), u_{2}^{\prime}, y_{1}, y_{2}) }{ P_{Y_{1}, Y_{2}}(y_{1}, y_{2}) }
\\
& \overset{\eqref{eq:ind}}{=}
\frac{ \sum_{u_{2}^{\prime} \in \mathbb{Z}_{q}} P_{X_{1}, Y_{1}}( f_{\gamma}( u_{1}, u_{2}^{\prime} ), y_{1} ) P_{X_{2}, Y_{2}}(u_{2}^{\prime}, y_{2}) }{ P_{Y_{1}, Y_{2}}(y_{1}, y_{2}) }
\\
& \overset{\eqref{eq:outputs}}{=}
\frac{ \sum_{u_{2}^{\prime} \in \mathbb{Z}_{q}} P_{X_{1}, Y_{1}}( f_{\gamma}( u_{1}, u_{2}^{\prime} ), y_{1} ) P_{X_{2}, Y_{2}}(u_{2}^{\prime}, y_{2}) }{ P_{Y_{1}}(y_{1}) P_{Y_{2}}(y_{2}) }
\\
& =
\sum_{u_{2}^{\prime} \in \mathbb{Z}_{q}} P_{X_{1}|Y_{1}}( f_{\gamma}( u_{1}, u_{2}^{\prime} ) \mid y_{1} ) P_{X_{2}|Y_{2}}(u_{2}^{\prime} \mid y_{2})
\label{eq:u1_y1y2}
\end{align}
for $(u_{1}, y_{1}, y_{2}) \in \mathbb{Z}_{q} \times \mathcal{Y}^{2}$.
Since
\begin{align}
P_{X_{1}|Y_{1}}( f_{\gamma}( u_{1}, u_{2}^{\prime} ) \mid [ r_{1} ]_{d_{1}} )
\overset{\eqref{eq:posteriori}}{=}
\begin{cases}
d_{1} / q
& \mathrm{if} \ u_{1} + \gamma \, u_{2}^{\prime} \equiv r_{1} \pmod{d_{1}} , \\
0
& \mathrm{otherwise}
\end{cases}
\end{align}
for $(u_{1}, u_{2}^{\prime}, [r_{1}]_{d_{1}}) \in \mathbb{Z}_{q}^{2} \times \mathcal{Y}$, and
\begin{align}
P_{X_{2}|Y_{2}}(u_{2}^{\prime} \mid [ r_{2} ]_{d_{2}})
\overset{\eqref{eq:posteriori}}{=}
\begin{cases}
d_{2} / q
& \mathrm{if} \ u_{2}^{\prime} \equiv r_{2} \pmod{d_{2}} , \\
0
& \mathrm{otherwise}
\end{cases}
\end{align}
for $(u_{2}^{\prime}, [r_{2}]_{d_{2}}) \in \mathbb{Z}_{q} \times \mathcal{Y}$, a term of the summation of the right-hand side of \eqref{eq:u1_y1y2} satisfies
\begin{align}
P_{X_{1}|Y_{1}}( f_{\gamma}( u_{1}, u_{2}^{\prime} ) \mid [ r_{1} ]_{d_{1}} ) P_{X_{2}|Y_{2}}(u_{2}^{\prime} \mid [ r_{2} ]_{d_{2}})
& =
\begin{cases}
\dfrac{ d_{1} \cdot d_{2} }{ q^{2} } 
& \mathrm{if} \ u_{1} + \gamma \, u_{2}^{\prime} \equiv r_{1} \pmod{d_{1}} , \\
& \hspace{3em} \ \: \, u_{2}^{\prime} \equiv r_{2} \pmod{d_{2}} , \\
0
& \mathrm{otherwise}
\end{cases}
\label{eq:u1_y1y2_part}
\end{align}
for $( u_{1}, u_{2}^{\prime}, [r_{1}]_{d_{1}}, [r_{2}]_{d_{2}} ) \in \mathbb{Z}_{q}^{2} \times \mathcal{Y}^{2}$.
In \eqref{eq:u1_y1y2_part}, the system of two congruences
\begin{align}
u_{1} + \gamma \, u_{2}^{\prime}
& \equiv
r_{1} \pmod{d_{1}} ,
\\
u_{2}^{\prime}
& \equiv
r_{2} \pmod{d_{2}}
\end{align}
can be rewritten as
\begin{align}
\gamma^{-1}( r_{1} - u_{1} )
& \equiv
u_{2}^{\prime} \pmod{d_{1}} ,
\label{eq:cong_u2p_1} \\
r_{2}
& \equiv
u_{2}^{\prime} \pmod{d_{2}} ,
\label{eq:cong_u2p_2}
\end{align}
where note that, for any $d \in \mathcal{D}_{q}$ and $\gamma \in \mathbb{Z}_{q}^{\times}$, there exists a unique modular multiplicative inverse $\gamma^{-1} \in \mathbb{Z}_{d}$ of $\gamma$, i.e., $\gamma^{-1} \gamma \equiv 1 \pmod{d}$, since $\gcd(\gamma, d) = 1$ for each $d \in \mathcal{D}_{q}$ and $\gamma \in \mathbb{Z}_{q}^{\times}$.
Then, \lemref{lem:chinese} shows that the system of two congruences \eqref{eq:cong_u2p_1} and \eqref{eq:cong_u2p_2} has a unique solution $u_{2}^{\prime} \in \mathbb{Z}_{q}$ modulo $\lcm( d_{1}, d_{2} )$ if and only if
\begin{align}
\gamma^{-1}( r_{1} - u_{1} )
\equiv
r_{2}
\pmod{ \gcd( d_{1}, d_{2} ) } .
\label{eq:cong_cond}
\end{align}
Thus for any $\gamma \in \mathbb{Z}_{q}^{\times}$ and $(u_{1}, [r_{1}]_{d_{1}}, [r_{2}]_{d_{2}}) \in \mathbb{Z}_{q} \times \mathcal{Y}^{2}$, if the congruence \eqref{eq:cong_cond} holds, there exists a unique $r^{\prime} \in \mathbb{Z}_{\lcm( d_{1}, d_{2} )}$ such that
\begin{align}
P_{X_{1}|Y_{1}}( f_{\gamma}( u_{1}, u_{2}^{\prime} ) \mid [ r_{1} ]_{d_{1}} ) P_{X_{2}|Y_{2}}(u_{2}^{\prime} \mid [ r_{2} ]_{d_{2}})
=
\begin{cases}
\dfrac{ d_{1} \cdot d_{2} }{ q^{2} } 
& \mathrm{if} \ r^{\prime} \equiv u_{2}^{\prime} \pmod{ \lcm( d_{1}, d_{2} ) } , \\
0
& \mathrm{otherwise}
\end{cases}
\label{eq:u1_y1y2_part2}
\end{align}
for $u_{2}^{\prime} \in \mathbb{Z}_{q}$;
and hence, we get
\begin{align}
P_{U_{1}|Y_{1}, Y_{2}}( u_{1} \mid [r_{1}]_{d_{1}}, [r_{2}]_{d_{2}} )
& \overset{\eqref{eq:u1_y1y2}}{=}
\sum_{u_{2}^{\prime} \in \mathbb{Z}_{q}} P_{X_{1}|Y_{1}}( f_{\gamma}(u_{1}, u_{2}^{\prime}) \mid [r_{1}]_{d_{1}} ) P_{X_{2}|Y_{2}}( u_{2}^{\prime} \mid [r_{2}]_{d_{2}} )
\\
& \overset{\eqref{eq:u1_y1y2_part2}}{=}
\sum_{u_{2}^{\prime} \in \mathbb{Z}_{q} \cap [ r^{\prime} ]_{\lcm( d_{1}, d_{2} )}} \frac{ d_{1} \cdot d_{2} }{ q^{2} }
\\
& \overset{\text{(a)}}{=}
\frac{ q }{ \lcm( d_{1}, d_{2} ) } \times \frac{ d_{1} \cdot d_{2} }{ q^{2} }
\\
& \overset{\text{(b)}}{=}
\frac{ \gcd( d_{1}, d_{2} ) }{ q }
\label{eq:u1_y1y2_1}
\end{align}
for $(u_{1}, [r_{1}]_{d_{1}}, [r_{2}]_{d_{2}}) \in \mathbb{Z}_{q} \times \mathcal{Y}^{2}$, where (a) follows by Lagrange's theorem and (b) follows by the identity
\begin{align}
\gcd(d_{1}, d_{2}) \cdot \lcm(d_{1}, d_{2})
=
d_{1} \cdot d_{2} .
\label{eq:gcd_lcm}
\end{align}
On the other hand, if the congruence \eqref{eq:cong_cond} does not hold, then \lemref{lem:chinese} shows
\begin{align}
P_{X_{1}|Y_{1}}( f_{\gamma}( u_{1}, u_{2}^{\prime} ) \mid [ r_{1} ]_{d_{1}} ) P_{X_{2}|Y_{2}}(u_{2}^{\prime} \mid [ r_{2} ]_{d_{2}})
=
0
\label{eq:u1_y1y2_part_NOT}
\end{align}
for $( u_{1}, u_{2}^{\prime}, [r_{1}]_{d_{1}}, [r_{2}]_{d_{2}} ) \in \mathbb{Z}_{q}^{2} \times \mathcal{Y}^{2}$.
Therefore, combining \eqref{eq:u1_y1y2_1} and \eqref{eq:u1_y1y2_part_NOT}, we have
\begin{align}
P_{U_{1}|Y_{1}, Y_{2}}( u_{1} \mid [r_{1}]_{d_{1}}, [r_{2}]_{d_{2}} )
=
\begin{cases}
\dfrac{ \gcd( d_{1}, d_{2} ) }{ q }
& \mathrm{if} \ r_{1} - \gamma \, r_{2} \equiv u_{1} \pmod{ \gcd( d_{1}, d_{2} ) } , \\
0
& \mathrm{otherwise}
\end{cases}
\label{eq:u1_y1y2_final}
\end{align}
for $(u_{1}, [r_{1}]_{d_{1}}, [r_{2}]_{d_{2}}) \in \mathbb{Z}_{q} \times \mathcal{Y}^{2}$, where note that the congruence \eqref{eq:cong_cond} is equivalent to
\begin{align}
r_{1} - \gamma \, r_{2}
\equiv
u_{1}
\pmod{ \gcd( d_{1}, d_{2} ) } .
\label{eq:cong_cond_2}
\end{align}
Then, for the a posteriori probability distribution \eqref{eq:u1_y1y2_final}, we observe that
\begin{align}
\mathcal{B} \big( ([r_{1}]_{d_{1}}, [r_{2}]_{d_{2}}) \big)
=
\left\{ ([r_{1}^{\prime}]_{d_{1}^{\prime}}, [r_{2}^{\prime}]_{d_{2}^{\prime}}) \in \mathcal{Y}^{2} \ \middle|
\begin{array}{l}
\gcd( d_{1}^{\prime}, d_{2}^{\prime} ) = \gcd( d_{1}, d_{2} ) , \\
r_{1}^{\prime} - \gamma \, r_{2}^{\prime} \equiv r_{1} - \gamma \, r_{2} \pmod{ \gcd( d_{1}, d_{2} ) }
\end{array}
\right\}
\label{eq:bin_gcd}
\end{align}
for $([r_{1}]_{d_{1}}, [r_{2}]_{d_{2}}) \in \mathcal{Y}^{2}$, where $\mathcal{B}( \cdot )$ is defined in \eqref{def:bin}.
Note that the set $\mathcal{B} \big( ([r_{1}]_{d_{1}}, [r_{2}]_{d_{2}}) \big)$ of \eqref{eq:bin_gcd} is identical to the set
\begin{align}
\mathcal{B}^{\prime}( [r]_{d} )
=
\left\{ ([r_{1}^{\prime}]_{d_{1}^{\prime}}, [r_{2}^{\prime}]_{d_{2}^{\prime}}) \in \mathcal{Y}^{2} \ \middle|
\begin{array}{l}
\gcd( d_{1}^{\prime}, d_{2}^{\prime} ) = d , \\
r_{1}^{\prime} - \gamma \, r_{2}^{\prime} \equiv r \pmod{ d }
\end{array}
\right\}
\label{eq:bin_gcd_prime}
\end{align}
for $[r]_{d} \in \mathcal{Y}$ when $d = \gcd( d_{1}, d_{2} )$ and $r \equiv r_{1} - \gamma \, r_{2} \pmod{d}$.
Moreover, the channel $V_{\ast}^{0} : \mathbb{Z}_{q} \to (\mathcal{Y}^{2})_{\ast}$ with respect to $V^{0} : \mathbb{Z}_{q} \to \mathcal{Y}^{2}$ is given by
\begin{align}
V_{\ast}^{0}( \mathcal{B}^{\prime}( [r]_{d} ) \mid u_{1} )
& \overset{\eqref{def:W_ast}}{=}
\sum_{(y_{1}, y_{2}) \in \mathcal{B}^{\prime}( [r]_{d} ) } V^{0}( y_{1}, y_{2} \mid u_{1} )
\\
& \overset{\eqref{eq:bin_gcd_prime}}{=}
\sum_{\substack{ d_{1}, d_{2} \in \mathcal{D}_{q} : \\ \gcd(d_{1}, d_{2}) = d }} \ \sum_{\substack{ (r_{1}, r_{2}) \in \mathbb{Z}_{d_{1}} \times \mathbb{Z}_{d_{2}} : \\ r_{1} - \gamma r_{2} \equiv r \!\!\!\pmod{d} }} V^{0}( [r_{1}]_{d_{1}}, [r_{2}]_{d_{2}} \mid u_{1} )
\\
& \overset{\eqref{eq:V0}}{=}
\sum_{\substack{ d_{1}, d_{2} \in \mathcal{D}_{q} : \\ \gcd(d_{1}, d_{2}) = d }} \ \sum_{\substack{ (r_{1}, r_{2}) \in \mathbb{Z}_{d_{1}} \times \mathbb{Z}_{d_{2}} : \\ r_{1} - \gamma r_{2} \equiv r \!\!\!\pmod{d} }} P_{Y_{1}, Y_{2}|U_{1}}( [r_{1}]_{d_{1}}, [r_{2}]_{d_{2}} \mid u_{1} )
\\
& =
\sum_{\substack{ d_{1}, d_{2} \in \mathcal{D}_{q} : \\ \gcd(d_{1}, d_{2}) = d }} \ \sum_{\substack{ (r_{1}, r_{2}) \in \mathbb{Z}_{d_{1}} \times \mathbb{Z}_{d_{2}} : \\ r_{1} - \gamma r_{2} \equiv r \!\!\!\pmod{d} }} \frac{ P_{U_{1}, Y_{1}, Y_{2}}( u_{1}, [r_{1}]_{d_{1}}, [r_{2}]_{d_{2}} ) }{ P_{U_{1}}( u_{1} ) }
\\
& =
\sum_{\substack{ d_{1}, d_{2} \in \mathcal{D}_{q} : \\ \gcd(d_{1}, d_{2}) = d }} \ \sum_{\substack{ (r_{1}, r_{2}) \in \mathbb{Z}_{d_{1}} \times \mathbb{Z}_{d_{2}} : \\ r_{1} - \gamma r_{2} \equiv r \!\!\!\pmod{d} }} \frac{ P_{Y_{1}, Y_{2}}( [r_{1}]_{d_{1}}, [r_{2}]_{d_{2}} ) P_{U_{1}|Y_{1}, Y_{2}}(u_{1} \mid [r_{1}]_{d_{1}}, [r_{2}]_{d_{2}} ) }{ (1/q) }
\\
& \overset{\eqref{eq:outputs}}{=}
\sum_{\substack{ d_{1}, d_{2} \in \mathcal{D}_{q} : \\ \gcd(d_{1}, d_{2}) = d }} \ \sum_{\substack{ (r_{1}, r_{2}) \in \mathbb{Z}_{d_{1}} \times \mathbb{Z}_{d_{2}} : \\ r_{1} - \gamma r_{2} \equiv r \!\!\!\pmod{d} }} q \times P_{Y_{1}}( [r_{1}]_{d_{1}} ) \times P_{Y_{2}}( [r_{2}]_{d_{2}} ) \times P_{U_{1}|Y_{1}, Y_{2}}(u_{1} \mid [r_{1}]_{d_{1}}, [r_{2}]_{d_{2}} ) 
\\
& \overset{\eqref{eq:marginal_out}}{=}
\sum_{\substack{ d_{1}, d_{2} \in \mathcal{D}_{q} : \\ \gcd(d_{1}, d_{2}) = d }} \ \sum_{\substack{ (r_{1}, r_{2}) \in \mathbb{Z}_{d_{1}} \times \mathbb{Z}_{d_{2}} : \\ r_{1} - \gamma r_{2} \equiv r \!\!\!\pmod{d} }} q \times \frac{ \varepsilon_{d_{1}} }{ d_{1} } \times \frac{ \varepsilon_{d_{2}} }{ d_{2} } \times P_{U_{1}|Y_{1}, Y_{2}}(u_{1} \mid [r_{1}]_{d_{1}}, [r_{2}]_{d_{2}} )
\\
& \overset{\eqref{eq:u1_y1y2_final}}{=}
\begin{cases}
\displaystyle
\sum \limits_{\substack{ d_{1}, d_{2} \in \mathcal{D}_{q} : \\ \gcd(d_{1}, d_{2}) = d }} \ \sum \limits_{\substack{ (r_{1}, r_{2}) \in \mathbb{Z}_{d_{1}} \times \mathbb{Z}_{d_{2}} : \\ r_{1} - \gamma r_{2} \equiv r \!\!\!\pmod{d} }} \dfrac{ q \, \varepsilon_{d_{1}} \, \varepsilon_{d_{2}} }{ d_{1} \, d_{2} } \times \frac{ \gcd(d_{1}, d_{2} ) }{ q }
& \mathrm{if} \ r_{1} - \gamma \, r_{2} \equiv u_{1} \pmod{ \gcd(d_{1}, d_{2}) } , \\
0
& \mathrm{otherwise}
\end{cases}
\\
& =
\begin{cases}
\displaystyle
\sum \limits_{\substack{ d_{1}, d_{2} \in \mathcal{D}_{q} : \\ \gcd(d_{1}, d_{2}) = d }} \ \sum \limits_{\substack{ (r_{1}, r_{2}) \in \mathbb{Z}_{d_{1}} \times \mathbb{Z}_{d_{2}} : \\ r_{1} - \gamma r_{2} \equiv r \!\!\!\pmod{d} }} \dfrac{ \varepsilon_{d_{1}} \, \varepsilon_{d_{2}} }{ d_{1} \, d_{2} } \times \gcd(d_{1}, d_{2} )
& \mathrm{if} \ r \equiv u_{1} \pmod{ d } , \\
0
& \mathrm{otherwise}
\end{cases}
\\
& \overset{\text{(a)}}{=}
\begin{cases}
\displaystyle
\sum \limits_{\substack{ d_{1}, d_{2} \in \mathcal{D}_{q} : \\ \gcd(d_{1}, d_{2}) = d }} \dfrac{ d_{1} \, d_{2} }{ \gcd( d_{1}, d_{2} ) } \times \dfrac{ \varepsilon_{d_{1}} \, \varepsilon_{d_{2}} }{ d_{1} \, d_{2} } \times \gcd(d_{1}, d_{2} )
& \mathrm{if} \ r \equiv u_{1} \pmod{ d } , \\
0
& \mathrm{otherwise}
\end{cases}
\\
& =
\begin{cases}
\displaystyle
\sum \limits_{\substack{ d_{1}, d_{2} \in \mathcal{D}_{q} : \\ \gcd(d_{1}, d_{2}) = d }} \varepsilon_{d_{1}} \, \varepsilon_{d_{2}}
& \mathrm{if} \ [ r ]_{d} = [ u_{1} ]_{d} , \\
0
& \mathrm{otherwise}
\end{cases}
\label{eq:V0_ast}
\end{align}
for $( u_{1}, \mathcal{B}^{\prime}([r]_{d}) ) \in \mathbb{Z}_{q} \times (\mathcal{Y}^{2})_{\ast}$, where the output alphabet $(\mathcal{Y}^{2})_{\ast}$ is given by
\begin{align}
(\mathcal{Y}^{2})_{\ast}
& =
\{ \mathcal{B}^{\prime}( [r]_{d} ) \mid d \in \mathcal{D}_{q} \ \mathrm{and} \ r \in \mathbb{Z}_{d} \}
\end{align}
and (a) follows by Lagrange's theorem.
Since $(\mathcal{Y}^{2})_{\ast}$ is isomorphic to $\mathcal{Y}$ with the mapping $\mathcal{B}^{\prime}( \cdot )$, the right-hand side of \eqref{eq:V0_ast} implies that the channel $V_{\ast}^{0}$ is identical to the channel of \defref{def:V} with the probability vector $( \varepsilon_{d}^{-} : d \in \mathcal{D}_{q} )$ calculated by \eqref{eq:E+-};
and therefore, \lemref{lem:bin} proves \thref{th:V} with respect to the channel $V^{0}$.

\subsection{Second part: Channel $V^{1}$}

In the second part, we consider the channel $V^{1} : \mathbb{Z}_{q} \to \mathcal{Y}^{2} \times \mathbb{Z}_{q}$, which is generated by $V \overset{\gamma}{\mapsto} (V^{0}, V^{1})$, as
\begin{align}
V^{1}(y_{1}, y_{2}, u_{1} \mid u_{2})
& \overset{\eqref{def:W1}}{=}
\frac{1}{q} V(y_{1} \mid f_{\gamma}(u_{1}, u_{2})) V(y_{2} \mid u_{2})
\\
& \overset{\eqref{eq:unif}}{=}
P_{X_{1}}( f_{\gamma}(u_{1}, u_{2}) ) V(y_{1} \mid f_{\gamma}(u_{1}, u_{2})) V(y_{2} \mid u_{2})
\\
& \overset{\eqref{def:x1y1}}{=}
P_{X_{1}}( f_{\gamma}(u_{1}, u_{2}) ) P_{Y_{1}|X_{1}}(y_{1} \mid f_{\gamma}(u_{1}, u_{2})) V(y_{2} \mid u_{2})
\\
& \overset{\eqref{def:x2y2}}{=}
P_{X_{1}}( f_{\gamma}(u_{1}, u_{2}) ) P_{Y_{1}|X_{1}}(y_{1} \mid f_{\gamma}(u_{1}, u_{2})) P_{Y_{2}|X_{2}}(y_{2} \mid u_{2})
\\
& =
\frac{ P_{X_{1}}(f_{\gamma}(u_{1}, u_{2})) P_{X_{2}}( u_{2} ) P_{Y_{1}|X_{1}}(y_{1} \mid f_{\gamma}(u_{1}, u_{2})) P_{Y_{2}|X_{2}}(y_{2} \mid u_{2}) }{ P_{X_{2}}( u_{2} ) }
\\
& =
\frac{ P_{X_{1}, Y_{1}}(f_{\gamma}(u_{1}, u_{2}), y_{1}) P_{X_{2}, Y_{2}}(u_{2}, y_{2}) }{ P_{X_{2}}( u_{2} ) }
\\
& \overset{\eqref{eq:ind}}{=}
\frac{ P_{X_{1} X_{2}, Y_{1}, Y_{2}}(f_{\gamma}(u_{1}, u_{2}), u_{2}, y_{1}, y_{2}) }{ P_{X_{2}}( u_{2} ) }
\\
& \overset{\eqref{def:x1x2}}{=}
\frac{ P_{U_{1} U_{2}, Y_{1}, Y_{2}}(u_{1}, u_{2}, y_{1}, y_{2}) }{ P_{U_{2}}(u_{2}) }
\\
& =
P_{Y_{1}, Y_{2}, U_{1} | U_{2}}(y_{1}, y_{2}, u_{1} \mid u_{2})
\label{eq:V1}
\end{align}
for $(u_{1}, u_{2}, y_{1}, y_{2}) \in \mathbb{Z}_{q}^{2} \times \mathcal{Y}^{2}$.
From \eqref{eq:outputs}, \eqref{eq:marginal_out}, and \eqref{eq:u1_y1y2_final}, we get
\begin{align}
r_{1} - \gamma \, r_{2} \equiv u_{1} \pmod{ \gcd(d_{1}, d_{2}) }
\ \ \mathrm{and} \ \
\min\{ \varepsilon_{d_{1}}, \varepsilon_{d_{2}} \} > 0
\iff
P_{U_{1}, Y_{1}, Y_{2}}( u_{1}, [r_{1}]_{d_{1}}, [r_{2}]_{d_{2}} ) > 0
\label{iff:outputs_1}
\end{align}
for $( u_{1}, [r_{1}]_{d_{1}}, [r_{2}]_{d_{2}} ) \in \mathbb{Z}_{q} \times \mathcal{Y}^{2}$.
Thus, if the congruence \eqref{eq:cong_cond_2} does not hold, then $V^{1}(y_{1}, y_{2}, u_{1} \mid u_{2}) = 0$;
and therefore, we assume that the congruence \eqref{eq:cong_cond_2} holds henceforth in the proof.
With attention to \eqref{iff:outputs_1}, the a posteriori probability of \eqref{eq:V1} is given by
\begin{align}
P_{U_{2}|Y_{1}, Y_{2}, U_{1}}(u_{2} \mid y_{1}, y_{2}, u_{1})
& =
\frac{ P_{U_{1}, U_{2}, Y_{1}, Y_{2}}(u_{1}, u_{2}, y_{1}, y_{2}) }{ P_{U_{1}, Y_{1}, Y_{2}}(u_{1}, y_{1}, y_{2}) }
\\
& =
\frac{ P_{U_{1}, U_{2}, Y_{1}, Y_{2}}(u_{1}, u_{2}, y_{1}, y_{2}) }{ P_{Y_{1}, Y_{2}}(y_{1}, y_{2}) P_{U_{1} | Y_{1}, Y_{2}}(u_{1} \mid y_{1}, y_{2}) }
\\
& \overset{\eqref{def:x1x2}}{=}
\frac{ P_{X_{1}, X_{2}, Y_{1}, Y_{2}}(f_{\gamma}(u_{1}, u_{2}), u_{2}, y_{1}, y_{2}) }{ P_{Y_{1}, Y_{2}}(y_{1}, y_{2}) P_{U_{1} | Y_{1}, Y_{2}}(u_{1} \mid y_{1}, y_{2}) }
\\
& \overset{\eqref{eq:ind}}{=}
\frac{ P_{X_{1}, Y_{1}}( f_{\gamma}( u_{1}, u_{2} ), y_{1} ) P_{X_{2}, Y_{2}}(u_{2}, y_{2}) }{ P_{Y_{1}, Y_{2}}(y_{1}, y_{2}) P_{U_{1} | Y_{1}, Y_{2}}(u_{1} \mid y_{1}, y_{2}) }
\\
& \overset{\eqref{eq:outputs}}{=}
\frac{ P_{X_{1}, Y_{1}}( f_{\gamma}( u_{1}, u_{2} ), y_{1} ) P_{X_{2}, Y_{2}}(u_{2}, y_{2}) }{ P_{Y_{1}}(y_{1}) P_{Y_{2}}(y_{2}) P_{U_{1} | Y_{1}, Y_{2}}(u_{1} \mid y_{1}, y_{2}) }
\\
& =
\frac{ P_{X_{1}|Y_{1}}( f_{\gamma}( u_{1}, u_{2} ) \mid y_{1} ) P_{X_{2}|Y_{2}}(u_{2} \mid y_{2}) }{ P_{U_{1} | Y_{1}, Y_{2}}(u_{1} \mid y_{1}, y_{2}) }
\label{eq:u2_y1y2u1}
\end{align}
for $(u_{1}, u_{2}, y_{1}, y_{2}) \in \mathbb{Z}_{q}^{2} \times \mathcal{Y}^{2}$.
Moreover, for any $\gamma \in \mathbb{Z}_{q}^{\times}$ and $(u_{1}, [r_{1}]_{d_{1}}, [r_{2}]_{d_{2}}) \in \mathbb{Z}_{q} \times \mathcal{Y}^{2}$, there exists a unique $r^{\prime} \in \mathbb{Z}_{\lcm(d_{1}, d_{2})}$ such that
\begin{align}
P_{U_{2}|Y_{1}, Y_{2}, U_{1}}(u_{2} \mid [r_{1}]_{d_{1}}, [r_{2}]_{d_{2}}, u_{1})
& \overset{\eqref{eq:u2_y1y2u1}}{=}
\frac{ P_{X_{1}|Y_{1}}( f_{\gamma}( u_{1}, u_{2} ) \mid [r_{1}]_{d_{1}} ) P_{X_{2}|Y_{2}}(u_{2} \mid [r_{2}]_{d_{2}}) }{ P_{U_{1} | Y_{1}, Y_{2}}(u_{1} \mid [r_{1}]_{d_{1}}, [r_{2}]_{d_{2}}) }
\\
& \overset{\eqref{eq:u1_y1y2_1}}{=}
P_{X_{1}|Y_{1}}( f_{\gamma}( u_{1}, u_{2} ) \mid [r_{1}]_{d_{1}} ) P_{X_{2}|Y_{2}}(u_{2} \mid [r_{2}]_{d_{2}}) \times \bigg( \frac{ \gcd( d_{1}, d_{2} ) }{ q } \bigg)^{-1}
\\
& \overset{\eqref{eq:u1_y1y2_part2}}{=}
\begin{cases}
\dfrac{ d_{1} \cdot d_{2} }{ q^{2} } \times \dfrac{ q }{ \gcd( d_{1}, d_{2} ) } 
& \mathrm{if} \ r^{\prime} \equiv u_{2} \pmod{ \lcm( d_{1}, d_{2} ) } , \\
0
& \mathrm{otherwise}
\end{cases}
\\
& \overset{\eqref{eq:gcd_lcm}}{=}
\begin{cases}
\dfrac{ \lcm(d_{1}, d_{2}) }{ q } 
& \mathrm{if} \ r^{\prime} \equiv u_{2} \pmod{ \lcm( d_{1}, d_{2} ) } , \\
0
& \mathrm{otherwise}
\end{cases}
\label{eq:u2_y1y2u1_final}
\end{align}
for $u_{2} \in \mathbb{Z}_{q}$.
Then, for the a posteriori probability distribution \eqref{eq:u2_y1y2u1_final}, we observe that
\begin{align}
\mathcal{B} \big( ([r_{1}]_{d_{1}}, [r_{2}]_{d_{2}}, u_{1}) \big)
=
\left\{ ([r_{1}^{\prime}]_{d_{1}^{\prime}}, [r_{2}^{\prime}]_{d_{2}^{\prime}}, u_{1}^{\prime}) \in \mathcal{Y}^{2} \times \mathbb{Z}_{q} \ \middle|
\begin{array}{l}
\lcm( d_{1}^{\prime}, d_{2}^{\prime} ) = \lcm( d_{1}, d_{2} ) ,
\\
\exists! r \in \mathbb{Z}_{\lcm( d_{1}, d_{2} )} \ \mathrm{such} \ \mathrm{that}
\\
\gamma^{-1}( r_{1}^{\prime} - u_{1}^{\prime} ) \equiv \gamma^{-1}( r_{1} - u_{1} ) \equiv r
\pmod{d_{1}} ,
\\
\hspace{9em} \;
r_{2}^{\prime} \equiv r_{2} \equiv r
\pmod{d_{2}}
\end{array}
\right\}
\label{eq:bin_lcm}
\end{align}
for $([r_{1}]_{d_{1}}, [r_{2}]_{d_{2}}, u_{1}) \in \mathcal{Y}^{2} \times \mathbb{Z}_{q}$, where $\mathcal{B}( \cdot )$ is defined in \eqref{def:bin}.
Note that the set $\mathcal{B} \big( ([r_{1}]_{d_{1}}, [r_{2}]_{d_{2}}, u_{1}) \big)$ of \eqref{eq:bin_lcm} is identical to the set
\begin{align}
\mathcal{B}^{\prime\prime}( [r]_{d} )
=
\left\{ ([r_{1}]_{d_{1}}, [r_{2}]_{d_{2}}, u_{1}) \in \mathcal{Y}^{2} \times \mathbb{Z}_{q} \ \middle|
\begin{array}{l}
\lcm( d_{1}, d_{2} ) = d ,
\\
\gamma^{-1}( r_{1} - u_{1} ) \equiv r
\pmod{d_{1}} ,
\\
\hspace{4em} \:
r_{2} \equiv r
\pmod{d_{2}}
\end{array}
\right\}
\label{eq:bin_lcm_prime}
\end{align}
for $[r]_{d} \in \mathcal{Y}$ when $d = \lcm(d_{1}, d_{2})$, $\gamma^{-1}( r_{1} - u_{1} ) \equiv r \pmod{d_{1}}$, and $r_{2} \equiv r \pmod{d_{2}}$.
Moreover, the channel $V_{\ast}^{1} : \mathbb{Z}_{q} \to (\mathcal{Y}^{2} \times \mathbb{Z}_{q})_{\ast}$ with respect to $V^{1} : \mathbb{Z}_{q} \to \mathcal{Y}^{2} \times \mathbb{Z}_{q}$ is given by
\begin{align}
V_{\ast}^{1}( \mathcal{B}^{\prime\prime}( [r]_{d} ) \mid u_{2} )
& \overset{\eqref{def:W_ast}}{=}
\sum_{(y_{1}, y_{2}, u_{1}) \in \mathcal{B}^{\prime\prime}( [r]_{d} ) } V^{1}( y_{1}, y_{2}, u_{1} \mid u_{2} )
\\
& \overset{\eqref{eq:bin_lcm_prime}}{=}
\sum_{\substack{ d_{1}, d_{2} \in \mathcal{D}_{q} : \\ \lcm(d_{1}, d_{2}) = d }} \ \sum_{\substack{ (u_{1}, r_{1}, r_{2}) \in \mathbb{Z}_{q} \times \mathbb{Z}_{d_{1}} \times \mathbb{Z}_{d_{2}} : \\ \gamma^{-1} (r_{1} - u_{1}) \equiv r \!\!\!\pmod{d_{1}}, \\ \qquad \; \; \; r_{2} \equiv r \!\!\!\pmod{d_{2}} }}
V^{1}( [r_{1}]_{d_{1}}, [r_{2}]_{d_{2}}, u_{1} \mid u_{2} )
\\
& \overset{\eqref{eq:V1}}{=}
\sum_{\substack{ d_{1}, d_{2} \in \mathcal{D}_{q} : \\ \lcm(d_{1}, d_{2}) = d }} \ \sum_{\substack{ (u_{1}, r_{1}, r_{2}) \in \mathbb{Z}_{q} \times \mathbb{Z}_{d_{1}} \times \mathbb{Z}_{d_{2}} : \\ \gamma^{-1} (r_{1} - u_{1}) \equiv r \!\!\!\pmod{d_{1}}, \\ \qquad \; \; \; r_{2} \equiv r \!\!\!\pmod{d_{2}} }}
P_{Y_{1}, Y_{2}, U_{1}|U_{2}}( [r_{1}]_{d_{1}}, [r_{2}]_{d_{2}}, u_{1} \mid u_{2} )
\\
& =
\sum_{\substack{ d_{1}, d_{2} \in \mathcal{D}_{q} : \\ \lcm(d_{1}, d_{2}) = d }} \ \sum_{\substack{ (u_{1}, r_{1}, r_{2}) \in \mathbb{Z}_{q} \times \mathbb{Z}_{d_{1}} \times \mathbb{Z}_{d_{2}} : \\ \gamma^{-1} (r_{1} - u_{1}) \equiv r \!\!\!\pmod{d_{1}}, \\ \qquad \; \; \; r_{2} \equiv r \!\!\!\pmod{d_{2}} }}
\frac{ P_{U_{1}, U_{2}, Y_{1}, Y_{2}}( u_{1}, u_{2}, [r_{1}]_{d_{1}}, [r_{2}]_{d_{2}}) }{ P_{U_{2}}( u_{2} ) }
\\
& =
\sum_{\substack{ d_{1}, d_{2} \in \mathcal{D}_{q} : \\ \lcm(d_{1}, d_{2}) = d }} \ \sum_{\substack{ (u_{1}, r_{1}, r_{2}) \in \mathbb{Z}_{q} \times \mathbb{Z}_{d_{1}} \times \mathbb{Z}_{d_{2}} : \\ \gamma^{-1} (r_{1} - u_{1}) \equiv r \!\!\!\pmod{d_{1}}, \\ \qquad \; \; \; r_{2} \equiv r \!\!\!\pmod{d_{2}} }} 1
\notag \\
& \qquad \times
\frac{ P_{Y_{1}, Y_{2}}([r_{1}]_{d_{1}}, [r_{2}]_{d_{2}}) P_{U_{1} | Y_{1}, Y_{2}}(u_{1} \mid [r_{1}]_{d_{1}}, [r_{2}]_{d_{2}}) P_{U_{2} | U_{1}, Y_{1}, Y_{2}}(u_{2} \mid u_{1}, [r_{1}]_{d_{1}}, [r_{2}]_{d_{2}}) }{ (1/q) }
\\
& \overset{\eqref{eq:outputs}}{=}
\sum_{\substack{ d_{1}, d_{2} \in \mathcal{D}_{q} : \\ \lcm(d_{1}, d_{2}) = d }} \ \sum_{\substack{ (u_{1}, r_{1}, r_{2}) \in \mathbb{Z}_{q} \times \mathbb{Z}_{d_{1}} \times \mathbb{Z}_{d_{2}} : \\ \gamma^{-1} (r_{1} - u_{1}) \equiv r \!\!\!\pmod{d_{1}}, \\ \qquad \; \; \; r_{2} \equiv r \!\!\!\pmod{d_{2}} }} q \, P_{Y_{1}}([r_{1}]_{d_{1}}) P_{Y_{2}}([r_{2}]_{d_{2}})
\notag \\
& \qquad \qquad \qquad \qquad \qquad \times
P_{U_{1} | Y_{1}, Y_{2}}(u_{1} \mid [r_{1}]_{d_{1}}, [r_{2}]_{d_{2}}) P_{U_{2} | U_{1}, Y_{1}, Y_{2}}(u_{2} \mid u_{1}, [r_{1}]_{d_{1}}, [r_{2}]_{d_{2}})
\\
& \overset{\eqref{eq:marginal_out}}{=}
\sum_{\substack{ d_{1}, d_{2} \in \mathcal{D}_{q} : \\ \lcm(d_{1}, d_{2}) = d }} \ \sum_{\substack{ (u_{1}, r_{1}, r_{2}) \in \mathbb{Z}_{q} \times \mathbb{Z}_{d_{1}} \times \mathbb{Z}_{d_{2}} : \\ \gamma^{-1} (r_{1} - u_{1}) \equiv r \!\!\!\pmod{d_{1}}, \\ \qquad \; \; \; r_{2} \equiv r \!\!\!\pmod{d_{2}} }} q \times \frac{ \varepsilon_{d_{1}} }{ d_{1} } \times \frac{ \varepsilon_{d_{2}} }{ d_{2} }
\notag \\
& \qquad \qquad \qquad \qquad \qquad \times
P_{U_{1} | Y_{1}, Y_{2}}(u_{1} \mid [r_{1}]_{d_{1}}, [r_{2}]_{d_{2}}) P_{U_{2} | U_{1}, Y_{1}, Y_{2}}(u_{2} \mid u_{1}, [r_{1}]_{d_{1}}, [r_{2}]_{d_{2}})
\\
& \overset{\eqref{eq:u1_y1y2_1}}{=}
\sum_{\substack{ d_{1}, d_{2} \in \mathcal{D}_{q} : \\ \lcm(d_{1}, d_{2}) = d }} \ \sum_{\substack{ (u_{1}, r_{1}, r_{2}) \in \mathbb{Z}_{q} \times \mathbb{Z}_{d_{1}} \times \mathbb{Z}_{d_{2}} : \\ \gamma^{-1} (r_{1} - u_{1}) \equiv r \!\!\!\pmod{d_{1}}, \\ \qquad \; \; \; r_{2} \equiv r \!\!\!\pmod{d_{2}} }} \frac{ q \, \varepsilon_{d_{1}} \, \varepsilon_{d_{2}} }{ d_{1} \, d_{2} } \times \frac{ \gcd(d_{1}, d_{2}) }{ q } \times P_{U_{2} | U_{1}, Y_{1}, Y_{2}}(u_{2} \mid u_{1}, [r_{1}]_{d_{1}}, [r_{2}]_{d_{2}})
\\
& \overset{\eqref{eq:gcd_lcm}}{=}
\sum_{\substack{ d_{1}, d_{2} \in \mathcal{D}_{q} : \\ \lcm(d_{1}, d_{2}) = d }} \ \sum_{\substack{ (u_{1}, r_{1}, r_{2}) \in \mathbb{Z}_{q} \times \mathbb{Z}_{d_{1}} \times \mathbb{Z}_{d_{2}} : \\ \gamma^{-1} (r_{1} - u_{1}) \equiv r \!\!\!\pmod{d_{1}}, \\ \qquad \; \; \; r_{2} \equiv r \!\!\!\pmod{d_{2}} }} \frac{ \varepsilon_{d_{1}} \, \varepsilon_{d_{2}} }{ \lcm(d_{1}, d_{2}) }  \times P_{U_{2} | U_{1}, Y_{1}, Y_{2}}(u_{2} \mid u_{1}, [r_{1}]_{d_{1}}, [r_{2}]_{d_{2}})
\\
& \overset{\eqref{eq:u2_y1y2u1_final}}{=}
\begin{cases}
\displaystyle
\sum_{\substack{ d_{1}, d_{2} \in \mathcal{D}_{q} : \\ \lcm(d_{1}, d_{2}) = d }} \ \sum_{\substack{ (u_{1}, r_{1}, r_{2}) \in \mathbb{Z}_{q} \times \mathbb{Z}_{d_{1}} \times \mathbb{Z}_{d_{2}} : \\ \gamma^{-1} (r_{1} - u_{1}) \equiv r \!\!\!\pmod{d_{1}}, \\ \qquad \; \; \; r_{2} \equiv r \!\!\!\pmod{d_{2}} }} \frac{ \varepsilon_{d_{1}} \, \varepsilon_{d_{2}} }{ \lcm(d_{1}, d_{2}) } \times \dfrac{ \lcm(d_{1}, d_{2}) }{ q }
& \mathrm{if} \ r \equiv u_{2} \pmod{ d } , \\
0
& \mathrm{otherwise}
\end{cases}
\\
& =
\begin{cases}
\displaystyle
\sum_{\substack{ d_{1}, d_{2} \in \mathcal{D}_{q} : \\ \lcm(d_{1}, d_{2}) = d }} \ \sum_{\substack{ (u_{1}, r_{1}, r_{2}) \in \mathbb{Z}_{q} \times \mathbb{Z}_{d_{1}} \times \mathbb{Z}_{d_{2}} : \\ \gamma^{-1} (r_{1} - u_{1}) \equiv r \!\!\!\pmod{d_{1}}, \\ \qquad \; \; \; r_{2} \equiv r \!\!\!\pmod{d_{2}} }} \frac{ \varepsilon_{d_{1}} \, \varepsilon_{d_{2}} }{ q }
& \mathrm{if} \ [r]_{d} = [u_{2}]_{d} , \\
0
& \mathrm{otherwise}
\end{cases}
\\
& \overset{\text{(a)}}{=}
\begin{cases}
\displaystyle
\sum_{\substack{ d_{1}, d_{2} \in \mathcal{D}_{q} : \\ \lcm(d_{1}, d_{2}) = d }} q \times \frac{ \varepsilon_{d_{1}} \, \varepsilon_{d_{2}} }{ q }
& \mathrm{if} \ [r]_{d} = [u_{2}]_{d} , \\
0
& \mathrm{otherwise}
\end{cases}
\\
& =
\begin{cases}
\displaystyle
\sum_{\substack{ d_{1}, d_{2} \in \mathcal{D}_{q} : \\ \lcm(d_{1}, d_{2}) = d }} \varepsilon_{d_{1}} \, \varepsilon_{d_{2}}
& \mathrm{if} \ [r]_{d} = [u_{2}]_{d} , \\
0
& \mathrm{otherwise}
\end{cases}
\label{eq:V1_ast}
\end{align}
for $(u_{2}, \mathcal{B}^{\prime\prime}( [r]_{d} )) \in \mathbb{Z}_{q} \times (\mathcal{Y}^{2} \times \mathbb{Z}_{q})_{\ast}$, where the output alphabet $(\mathcal{Y}^{2} \times \mathbb{Z}_{q})_{\ast}$ is given by
\begin{align}
(\mathcal{Y}^{2} \times \mathbb{Z}_{q})_{\ast}
=
\{ \mathcal{B}^{\prime\prime}( [r]_{d} ) \mid d \in \mathcal{D}_{q} \ \mathrm{and} \ r \in \mathbb{Z}_{d} \}
\end{align}
and (a) follows by Lagrange's theorem.
Since $(\mathcal{Y}^{2} \times \mathbb{Z}_{q})_{\ast}$ is isomorphic to $\mathcal{Y}$ with the mapping $\mathcal{B}^{\prime\prime}( \cdot )$, the right-hand side of \eqref{eq:V1_ast} implies that the channel $V_{\ast}^{1}$ is identical to the channel of \defref{def:V} with the probability vector $( \varepsilon_{d}^{+} : d \in \mathcal{D}_{q} )$ calculated by \eqref{eq:E+-};
and therefore, \lemref{lem:bin} proves \thref{th:V} with respect to the channel $V^{1}$.
\end{IEEEproof}

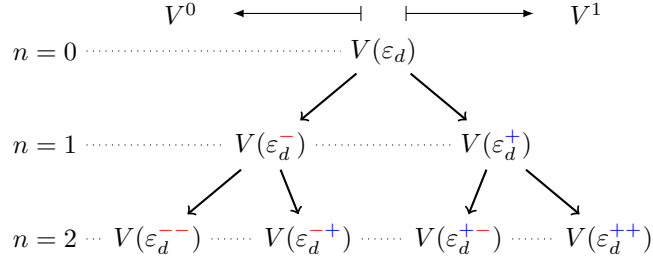
\begin{figure}[!t]
\centering
\begin{tikzpicture}
\node (v) at (0, 0) {$V( \varepsilon_{d} )$};
\node (v0) at (-1.5, -1.25) {$V( \varepsilon_{d}^{{\color{red} -}} )$};
\node (v1) at (1.5, -1.25) {$V( \varepsilon_{d}^{{\color{blue} +}} )$};
\node (v00) at (-3, -2.5) {$V( \varepsilon_{d}^{{\color{red} -}{\color{red} -}} )$};
\node (v01) at (-1, -2.5) {$V( \varepsilon_{d}^{{\color{red} -}{\color{blue} +}} )$};
\node (v10) at (1, -2.5) {$V( \varepsilon_{d}^{{\color{blue} +}{\color{red} -}} )$};
\node (v11) at (3, -2.5) {$V( \varepsilon_{d}^{{\color{blue} +}{\color{blue} +}} )$};
\node (n0) at (-4.5, 0) {$n = 0$};
\node (n1) at (-4.5, -1.25) {$n = 1$};
\node (n2) at (-4.5, -2.5) {$n = 2$};
\draw [->, thick] (v) -- (v0);
\draw [->, thick] (v) -- (v1);
\draw [->, thick] (v0) -- (v00);
\draw [->, thick] (v0) -- (v01);
\draw [->, thick] (v1) -- (v10);
\draw [->, thick] (v1) -- (v11);
\draw [dotted] (n0) -- (v);
\draw [dotted] (n1) -- (v0) -- (v1);
\draw [dotted] (n2) -- (v00) -- (v01) -- (v10) -- (v11) ;
\draw [-latex] (-0.3, 0.5) -- (-2, 0.5) node [left = 1em] {$V^{0}$};
\draw [-latex] (0.3, 0.5) -- (2, 0.5) node [right = 1em] {$V^{1}$};
\draw (-0.3, 0.6) -- (-0.3, 0.4);
\draw (0.3, 0.6) -- (0.3, 0.4);
\end{tikzpicture}
\caption{Recursive calculation of the polar transformation of the channel $V$ (cf. \thref{th:V}).
For simplicity, the channel $V( \varepsilon_{d} )$ denotes $V_{(q)}( \varepsilon_{d} : d \in \mathcal{D}_{q} )$.}
\label{fig:V}
\end{figure}

\begin{figure}[!t]
\centering
\begin{overpic}[width = 1\hsize, clip]{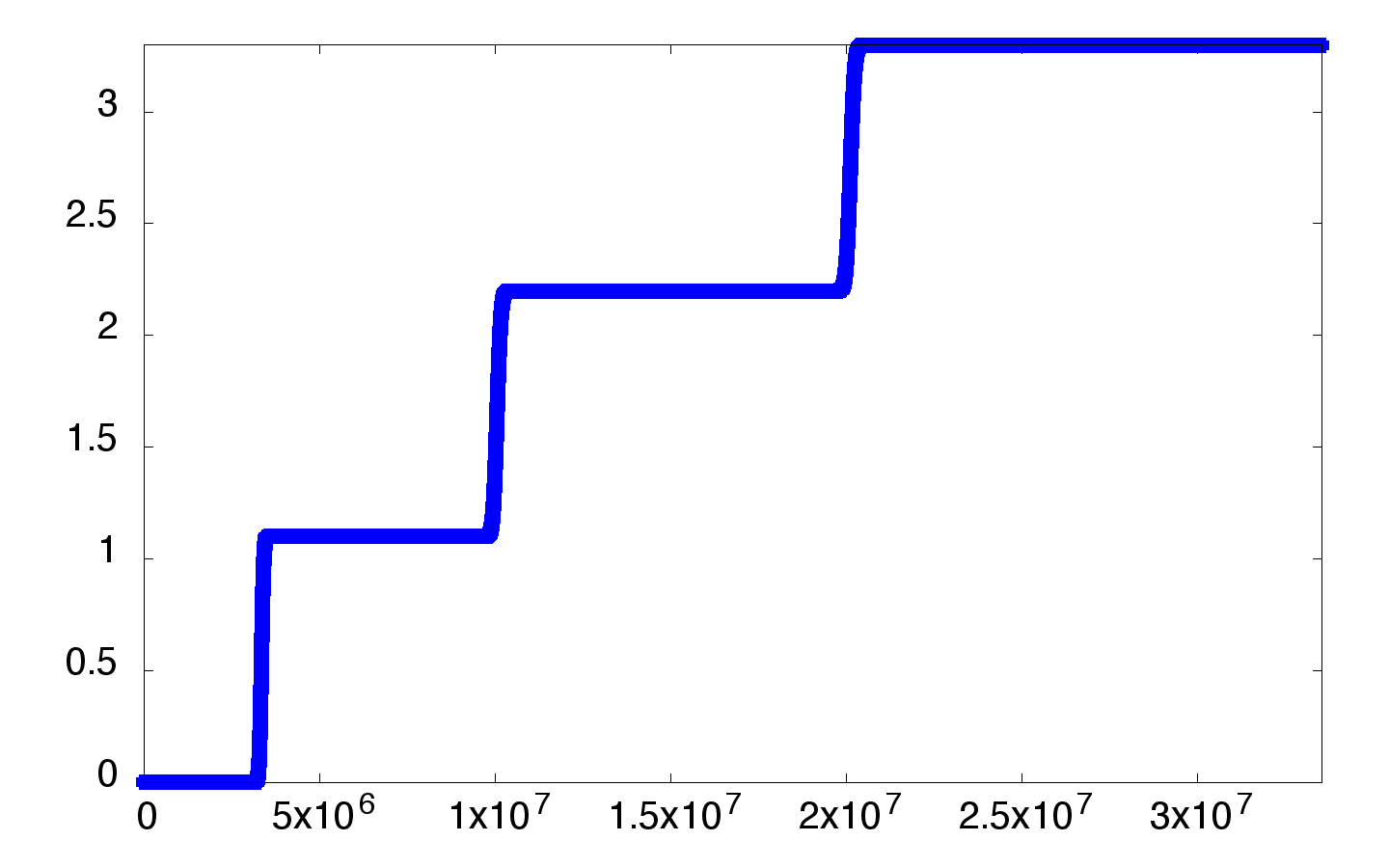}
\put(35, -2.5){index $i$ (in increasing order of $I \big( V_{2^{n}}^{(i)} \big)$)}
\put(0, 20){\rotatebox{90}{symmetric capacity $I \big( V_{2^{n}}^{(i)} \big)$}}
\put(2, 58){\scriptsize [nats]}
\put(69, 30){$I \big( V_{2^{n}}^{(i)} \big) \approx \ln 27$}
\put(72, 34){\vector(1, 3){8}}
\put(60, 15){$\approx \ln 9$}
\put(59, 16){\vector(-1, 3){7.75}}
\put(38, 10){$\approx \ln 3$}
\put(37, 11){\vector(-2, 3){7.5}}
\put(20, 33){$\approx 0$}
\put(19, 34){\vector(-1, -4){6.5}}
\end{overpic}
\hspace{0pt}
\caption{Multilevel polarization of the symmetric capacity $\{ I \big( V_{2^{n}}^{(i)} \big) \mid 0 \le i < 2^{n} \}$ of the channel $V_{(q)}( \varepsilon_{d} : d \in \mathcal{D}_{q} )$ with $q = 27 = 3^{3}$, $( \varepsilon_{d} : d \in \mathcal{D}_{27} ) = (\varepsilon_{1}, \varepsilon_{3}, \varepsilon_{9}, \varepsilon_{27}) = (1/10, 2/10, 3/10, 4/10)$, and $n = 25$.
The proportion of $I \big( V_{2^{n}}^{(i)} \big) \approx \ln d$ is nearly equal to $\varepsilon_{d}$ for each $d \in \mathcal{D}_{27}$ (cf.~\thref{th:martingale}).}
\label{fig:V27}
\end{figure}

\begin{figure}[!t]
\centering
\begin{overpic}[width = 1\hsize, clip]{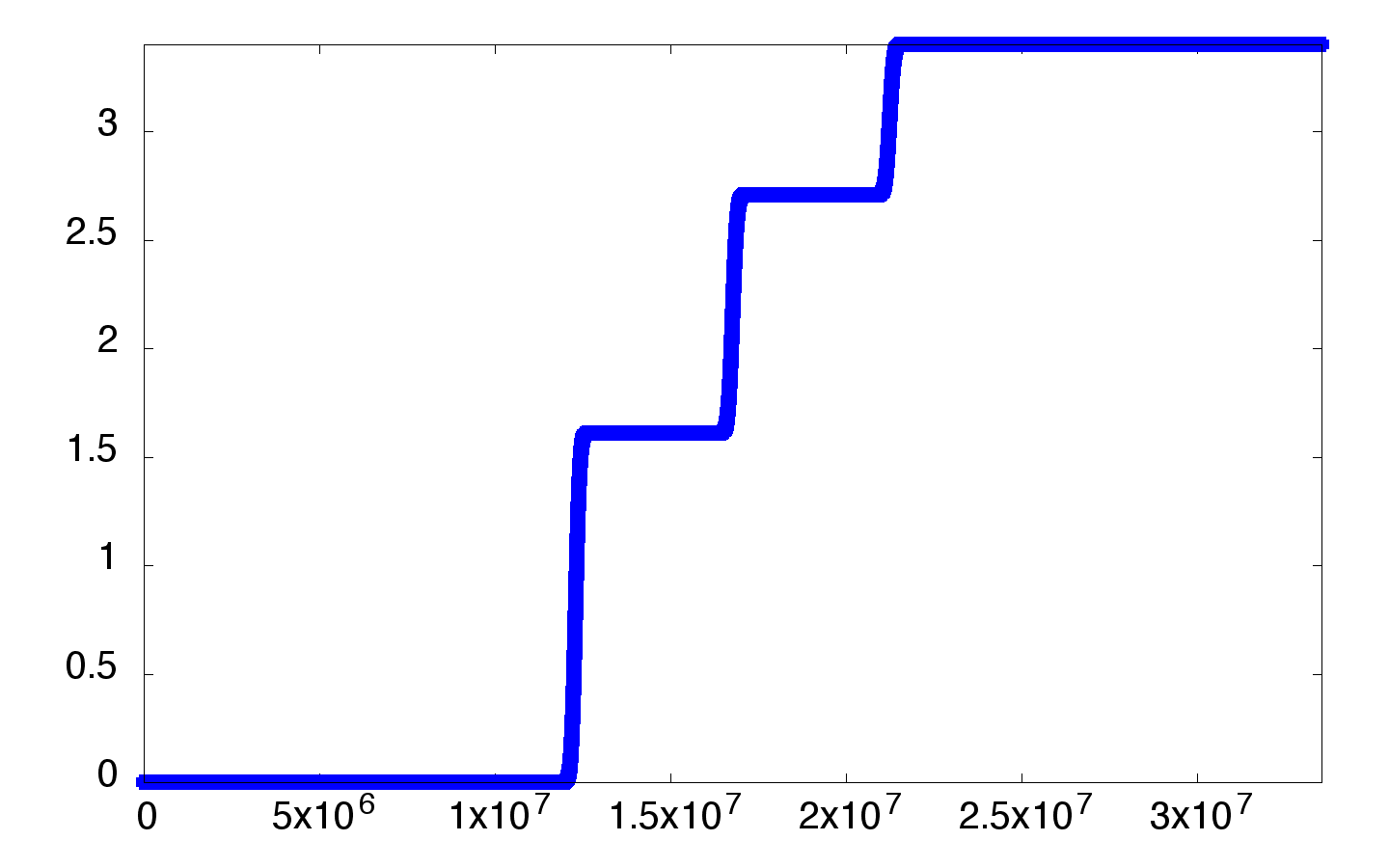}
\put(35, -2.5){index $i$ (in increasing order of $I \big( V_{2^{n}}^{(i)} \big)$)}
\put(0, 20){\rotatebox{90}{symmetric capacity $I \big( V_{2^{n}}^{(i)} \big)$}}
\put(2, 58){\scriptsize [nats]}
\put(18, 20){$I \big( V_{2^{n}}^{(i)} \big) \approx 0$}
\put(17, 21){\vector(-1, -4){3.5}}
\put(52, 15){$\approx \ln 5$}
\put(51, 16){\vector(-1, 4){3.25}}
\put(66, 30){$\approx \ln 15$}
\put(65, 31){\vector(-1, 3){5.3}}
\put(82, 43){$\approx \ln 30$}
\put(81, 44){\vector(-1, 4){3.5}}
\end{overpic}
\hspace{0pt}
\caption{Multilevel polarization of the symmetric capacity $\{ I \big( V_{2^{n}}^{(i)} \big) \mid 0 \le i < 2^{n} \}$ of the channel $V_{(q)}( \varepsilon_{d} : d \in \mathcal{D}_{q} )$ with $q = 30 = 2 \cdot 3 \cdot 5$, $( \varepsilon_{d} : d \in \mathcal{D}_{30} ) = (\varepsilon_{1}, \varepsilon_{2}, \varepsilon_{3}, \varepsilon_{5}, \varepsilon_{6}, \varepsilon_{10}, \varepsilon_{15}, \varepsilon_{30}) = (0, 3/30, 5/30, 7/30, 3/30, 5/30, 7/30, 0)$, and $n = 25$.}
\label{fig:V30}
\end{figure}
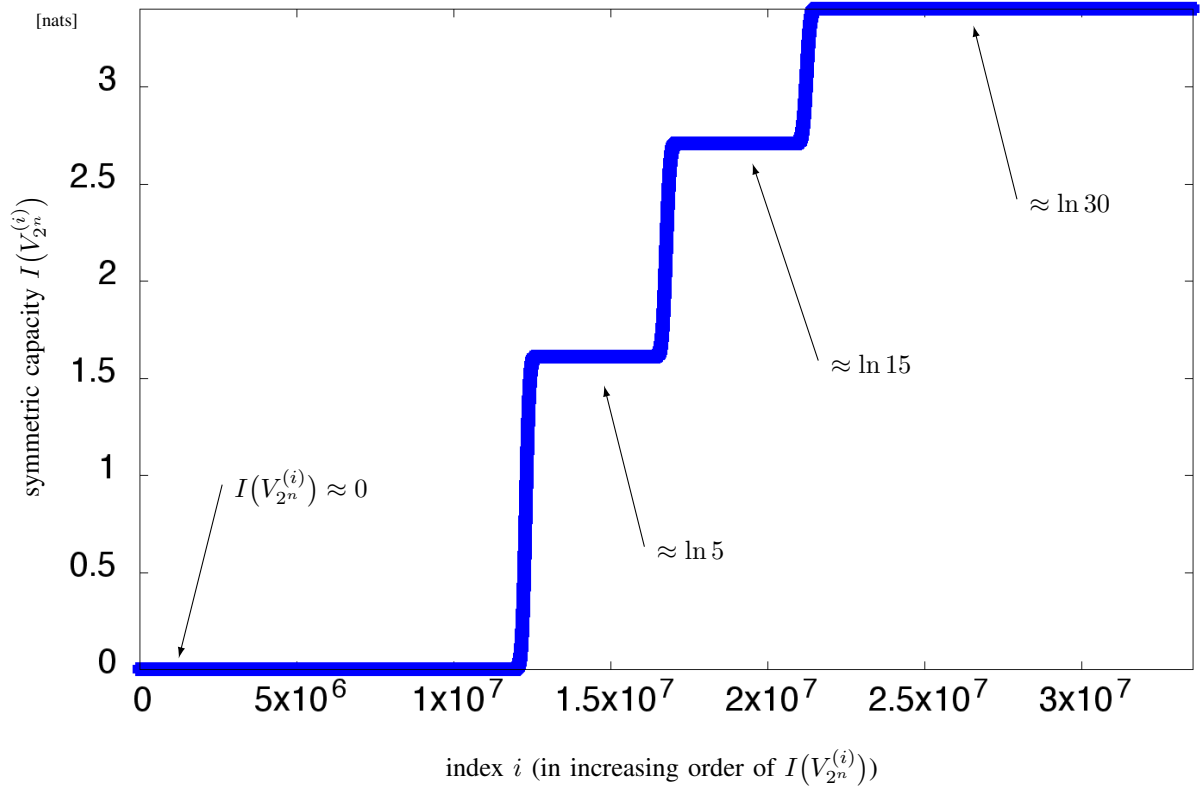

Since the polar transformation of the channel $V_{(q)}( \varepsilon_{d} : d \in \mathcal{D}_{q} )$ can be exactly approximated by other channels $V_{(q)}( \varepsilon_{d}^{-} : d \in \mathcal{D}_{q} )$ and $V_{(q)}( \varepsilon_{d}^{+} : d \in \mathcal{D}_{q} )$, it is enough to propagate the probability vector $( \varepsilon_{d} : d \in \mathcal{D}_{q} )$ recursively by using \eqref{eq:E+-}.
We illustrate this recursive calculation in \figref{fig:V}.
Using the recursive calculation, Figs.~\ref{fig:V27}~and~\ref{fig:V30} illustrate the multilevel polarizations of the (ordinary) symmetric capacity $I( V )$ of the channel $V$ with $q = 27$ and $q = 30$, respectively.

It is easy to verify that \thref{th:V} contains \propref{prop:BEC} as follows:
Since $V_{(2)}( \varepsilon_{1}, \varepsilon_{2} )$ is $\mathrm{BEC}( \varepsilon_{1} )$ and $\varepsilon_{2} = 1 - \varepsilon_{1}$, it follows from \thref{th:V} that 
\begin{align}
\varepsilon_{1}^{-}
& =
\sum_{d_{1}, d_{2} \in \mathcal{D}_{q} : \gcd(d_{1}, d_{2}) = 1} \varepsilon_{d_{1}} \varepsilon_{d_{2}}
\\
& =
\varepsilon_{1}^{2} + 2 \varepsilon_{1} \varepsilon_{2}
\\
& =
\varepsilon_{1}^{2} + 2 \varepsilon_{1} (1 - \varepsilon_{1})
\\
& =
2 \varepsilon_{1} - \varepsilon_{1}^{2} ,
\\
\varepsilon_{1}^{+}
& =
\sum_{d_{1}, d_{2} \in \mathcal{D}_{q} : \lcm(d_{1}, d_{2}) = 1} \varepsilon_{d_{1}} \varepsilon_{d_{2}}
\\
& =
\varepsilon_{1}^{2} ,
\end{align}
which are identical to \propref{prop:BEC}.

\begin{remark}
We now revisit the polar transformation of the channel $W$, defined in \cite[Fig.~4: Channel~2]{sahebi}.
When we use the pair of recursive formulas \cite[Eqs.~(3) and~(4)]{sahebi}, it can be confirmed that $I(W^{0}) + I(W^{1}) = 2 I(W)$ does not hold in general.
However, when we use the pair of recursive formulas \cite[Eq.~(3)]{sahebi_arxiv} and \cite[Eq.~(4)]{sahebi}, the identity $I(W^{0}) + I(W^{1}) = 2 I(W)$ holds.
Then, it can be verified that \thref{th:V} yields the pair of recursive formulas \cite[Eq.~(3)]{sahebi_arxiv} and \cite[Eq.~(4)]{sahebi}.
\end{remark}

\subsection{Special cases: the input alphabet size is a prime power}
\label{sect:prime_power}

In this subsection, we consider the polar transformation of the channel $V$ when the input alphabet size $q$ is a prime power, i.e., $q = p^{m}$ for some prime $p$ and some $m \in \mathbb{N}$.
The following corollary is directly derived from \thref{th:V}.

\begin{corollary}
\label{cor:V}
If the input alphabet size $q$ is a prime power, then \eqref{eq:E+-} can be rewritten as
\begin{align}
\varepsilon_{d}^{-}
& =
\varepsilon_{d} \cdot \Bigg( \varepsilon_{d} + 2 \sum_{d^{\prime} \in \mathcal{D}_{q} : d^{\prime} > d} \varepsilon_{d^{\prime}} \Bigg) ,
\label{eq:E-_primepower} \\
\varepsilon_{d}^{+}
& =
\varepsilon_{d} \cdot \Bigg( \varepsilon_{d} + 2 \sum_{d^{\prime} \in \mathcal{D}_{q} : d^{\prime} < d} \varepsilon_{d^{\prime}} \Bigg) .
\label{eq:E+_primepower}
\end{align}
In addition, it holds that
\begin{align}
\varepsilon_{d}^{-} + \varepsilon_{d}^{+}
=
2 \, \varepsilon_{d}
\label{eq:conservation_eps}
\end{align}
for $d \in \mathcal{D}_{q}$.
\end{corollary}

\begin{IEEEproof}[Proof of \corref{cor:V}]
Suppose that $q = p^{m}$ for some prime $p$ and some $m \in \mathbb{N}$, i.e., the input alphabet size $q$ is a prime power.
Then, it is easy to see that $\mathcal{D}_{q} = \{ p^{i} \mid 0 \le i \le m \} = \{ 1, p, p^{2} \dots, p^{m} \}$.
Since $\gcd( p^{i}, p^{j} ) = p^{i}$ for $0 \le i \le j$, we get
\begin{align}
\{ (d_{1}, d_{2}) \in \mathcal{D}_{q}^{2} \mid \gcd(d_{1}, d_{2}) = d \}
=
\{ (d, d) \}
\cup
\{ (d_{1}, d) \mid d_{1} > d \}
\cup
\{ (d, d_{2}) \mid d_{2} > d \}
\label{eq:decompose_gcd}
\end{align}
for $d \in \mathcal{D}_{q}$;
and therefore, we have
\begin{align}
\varepsilon_{d}^{-}
& \overset{\eqref{eq:E+-}}{=}
\sum_{\substack{ d_{1}, d_{2} \in \mathcal{D}_{q} : \\ \gcd(d_{1}, d_{2}) = d }} \varepsilon_{d_{1}} \, \varepsilon_{d_{2}}
\\
& \overset{\eqref{eq:decompose_gcd}}{=}
\varepsilon_{d} \, \varepsilon_{d}
+
\sum_{d_{1} \in \mathcal{D}_{q} : d_{1} > d} \varepsilon_{d_{1}} \, \varepsilon_{d}
+
\sum_{d_{2} \in \mathcal{D}_{q} : d_{2} > d} \varepsilon_{d} \, \varepsilon_{d_{2}}
\\
& =
\varepsilon_{d}^{2} + 2 \sum_{d^{\prime} \in \mathcal{D}_{q} : d^{\prime} > d} \varepsilon_{d^{\prime}} \, \varepsilon_{d}
\\
& =
\varepsilon_{d} \cdot \Bigg( \varepsilon_{d} + 2 \sum_{d^{\prime} \in \mathcal{D}_{q} : d^{\prime} > d} \varepsilon_{d^{\prime}} \Bigg)
\end{align}
for $d \in \mathcal{D}_{q}$, which is \eqref{eq:E-_primepower}.

Similarly, since $\lcm(p^{i}, p^{j}) = p^{j}$ for $0 \le i \le j$, we get
\begin{align}
\{ (d_{1}, d_{2}) \in \mathcal{D}_{q}^{2} \mid \gcd(d_{1}, d_{2}) = d \}
=
\{ (d, d) \}
\cup
\{ (d_{1}, d) \mid d_{1} < d \}
\cup
\{ (d, d_{2}) \mid d_{2} < d \}
\label{eq:decompose_lcm}
\end{align}
for $d \in \mathcal{D}_{q}$;
and therefore, we have
\begin{align}
\varepsilon_{d}^{+}
& \overset{\eqref{eq:E+-}}{=}
\sum_{\substack{ d_{1}, d_{2} \in \mathcal{D}_{q} : \\ \lcm(d_{1}, d_{2}) = d }} \varepsilon_{d_{1}} \, \varepsilon_{d_{2}}
\\
& \overset{\eqref{eq:decompose_lcm}}{=}
\varepsilon_{d} \, \varepsilon_{d}
+
\sum_{d_{1} \in \mathcal{D}_{q} : d_{1} < d} \varepsilon_{d_{1}} \, \varepsilon_{d}
+
\sum_{d_{2} \in \mathcal{D}_{q} : d_{2} < d} \varepsilon_{d} \, \varepsilon_{d_{2}}
\\
& =
\varepsilon_{d}^{2} + 2 \sum_{d^{\prime} \in \mathcal{D}_{q} : d^{\prime} < d} \varepsilon_{d^{\prime}} \, \varepsilon_{d}
\\
& =
\varepsilon_{d} \cdot \Bigg( \varepsilon_{d} + 2 \sum_{d^{\prime} \in \mathcal{D}_{q} : d^{\prime} < d} \varepsilon_{d^{\prime}} \Bigg)
\end{align}
for $d \in \mathcal{D}_{q}$, which is \eqref{eq:E+_primepower}.

Finally, a simple calculation yields
\begin{align}
\varepsilon_{d}^{-} + \varepsilon_{d}^{+}
& \overset{\eqref{eq:E-_primepower}}{=}
\varepsilon_{d} \cdot \Bigg( \varepsilon_{d} + 2 \sum_{d^{\prime} \in \mathcal{D}_{q} : d^{\prime} > d} \varepsilon_{d^{\prime}} \Bigg)
+
\varepsilon_{d}^{+}
\\
& \overset{\eqref{eq:E+_primepower}}{=}
\varepsilon_{d} \cdot \Bigg( \varepsilon_{d} + 2 \sum_{d^{\prime} \in \mathcal{D}_{q} : d^{\prime} > d} \varepsilon_{d^{\prime}} \Bigg)
+
\varepsilon_{d} \cdot \Bigg( \varepsilon_{d} + 2 \sum_{d^{\prime} \in \mathcal{D}_{q} : d^{\prime} < d} \varepsilon_{d^{\prime}} \Bigg)
\\
& =
2 \, \varepsilon_{d} \, \varepsilon_{d} + 2 \, \varepsilon_{d} \cdot \Bigg( \sum_{d^{\prime} \in \mathcal{D}_{q} : d^{\prime} > d} \varepsilon_{d^{\prime}} \Bigg) + 2 \, \varepsilon_{d} \cdot \Bigg( \sum_{d^{\prime} \in \mathcal{D}_{q} : d^{\prime} < d} \varepsilon_{d^{\prime}} \Bigg)
\\
& =
2 \, \varepsilon_{d} \cdot \Bigg( \underbrace{ \sum_{d^{\prime} \in \mathcal{D}_{q}} \varepsilon_{d^{\prime}} }_{ = 1 } \Bigg)
\\
& =
2 \, \varepsilon_{d}
\end{align}
for $d \in \mathcal{D}_{q}$.
This completes the proof of \corref{cor:V}.
\end{IEEEproof}

In \corref{cor:V}, note that $\mathcal{D}_{q} = \{ 1, p, p^{2}, \dots, p^{m-1}, p^{m} \}$ when $q = p^{m}$ for some prime $p$ and some $m \in \mathbb{N}$.
\corref{cor:V} can be reduced to the results of \cite[Section~III]{park2} when the input alphabet size $q$ is a power of two.
Moreover, the following corollary also directly follows from \corref{cor:V}.

\begin{corollary}
\label{cor:ineq}
Consider the polar transformation $V \overset{\gamma}{\mapsto} (V^{0},V^{1})$ with $\gamma \in \mathbb{Z}_{q}^{\times}$.
If the input alphabet size $q$ is a prime power, then
\begin{align}
I_{\alpha}( V^{0} ) + I_{\alpha}( V^{1} )
\ge
2 I_{\alpha}( V )
\quad & \mathrm{for} \ 0 \le \alpha \le 1 ,
\label{eq:ineq_1} \\
I_{\alpha}( V^{0} ) + I_{\alpha}( V^{1} )
\le
2 I_{\alpha}( V )
\quad & \mathrm{for} \ 1 \le \alpha \le \infty .
\label{eq:ineq_2}
\end{align}
\end{corollary}

\begin{IEEEproof}[Proof of \corref{cor:ineq}]
A direct calculation shows
\begin{align}
I_{\alpha}( V )
& \overset{\eqref{def:symmetric_alpha}}{=}
\frac{ \alpha }{ \alpha-1 } \ln \Bigg[ \sum_{y \in \mathcal{Y}} \Bigg( \sum_{x \in \mathbb{Z}_{q}} \frac{1}{q} V(y \mid x)^{\alpha} \Bigg)^{1/\alpha} \Bigg]
\\
& \overset{\eqref{def:eq:V}}{=}
\frac{ \alpha }{ \alpha-1 } \ln \Bigg[ \sum_{[r]_{d} \in \mathcal{Y}} \Bigg( \sum_{x \in \mathbb{Z}_{q} \cap [r]_{d}} \frac{ 1 }{ q } \times \varepsilon_{d}^{\alpha} \Bigg)^{1/\alpha} \Bigg]
\\
& \overset{\text{(a)}}{=}
\frac{ \alpha }{ \alpha-1 } \ln \Bigg[ \sum_{[r]_{d} \in \mathcal{Y}} \Bigg( \frac{q}{d} \times \frac{ 1 }{ q } \times \varepsilon_{d}^{\alpha} \Bigg)^{1/\alpha} \Bigg]
\\
& =
\frac{ \alpha }{ \alpha-1 } \ln \Bigg[ \sum_{[r]_{d} \in \mathcal{Y}} \Bigg( \frac{1}{d} \times \varepsilon_{d}^{\alpha} \Bigg)^{1/\alpha} \Bigg]
\\
& =
\frac{ \alpha }{ \alpha-1 } \ln \Bigg[ \sum_{[r]_{d} \in \mathcal{Y}} \bigg( \frac{ 1 }{ d } \bigg)^{1/\alpha} \times \varepsilon_{d} \Bigg]
\\
& =
\frac{ \alpha }{ \alpha-1 } \ln \Bigg[ \sum_{d \in \mathcal{D}_{q}} \sum_{r \in \mathbb{Z}_{d}} \bigg( \frac{ 1 }{ d } \bigg)^{1/\alpha} \times \varepsilon_{d} \Bigg]
\\
& =
\frac{ \alpha }{ \alpha-1 } \ln \Bigg[ \sum_{d \in \mathcal{D}_{q}} d \times \bigg( \frac{ 1 }{ d } \bigg)^{1/\alpha} \times \varepsilon_{d} \Bigg]
\\
& =
\frac{ \alpha }{ \alpha-1 } \ln \Bigg[ \sum_{d \in \mathcal{D}_{q}} \varepsilon_{d} \cdot \Big( d^{(\alpha-1)/\alpha} \Big) \Bigg]
\label{eq:Ialpha_V}
\end{align}
for $\alpha \in (0, 1) \cup (1, \infty)$, where (a) follows by Lagrange's theorem.
In addition, for $\alpha \in \{ 0, 1, \infty \}$, we get
\begin{align}
I_{0}(V)
& \overset{\eqref{def:alpha_0}}{=}
\lim_{\alpha \to 0^{+}} I_{\alpha}( V )
\\
& \overset{\eqref{eq:Ialpha_V}}{=}
\lim_{\alpha \to 0^{+}} \frac{ \alpha }{ \alpha-1 } \ln \Bigg[ \sum_{d \in \mathcal{D}_{q}} \varepsilon_{d} \cdot \Big( d^{(\alpha-1)/\alpha} \Big) \Bigg]
\\
& =
\lim_{\alpha \to 0^{+}} \ln \Bigg[ \sum_{d \in \mathcal{D}_{q}} \varepsilon_{d} \cdot \Big( d^{(\alpha-1)/\alpha} \Big) \Bigg]^{\alpha/(\alpha-1)}
\\
& =
\lim_{\beta \to -\infty} \ln \Bigg[ \sum_{d \in \mathcal{D}_{q}} \varepsilon_{d} \cdot \Big( d^{\beta} \Big) \Bigg]^{1/\beta}
\\
& =
\min_{d \in \mathcal{D}_{q} : \varepsilon_{d} > 0} \Big( \ln d \Big) ,
\label{eq:I0_V} \\
I_{1}( V )
& \overset{\eqref{def:alpha_1}}{=}
\lim_{\alpha \to 1} I_{\alpha}( V )
\\
& \overset{\eqref{eq:Ialpha_V}}{=}
\lim_{\alpha \to 1} \frac{ \alpha }{ \alpha-1 } \ln \Bigg[ \sum_{d \in \mathcal{D}_{q}} \varepsilon_{d} \cdot \Big( d^{(\alpha-1)/\alpha} \Big) \Bigg]
\\
& =
\lim_{\alpha \to 1} \bigg( \frac{ \alpha-1 }{ \alpha } \bigg)^{-1} \ln \Bigg[ \sum_{d \in \mathcal{D}_{q}} \varepsilon_{d} \cdot \Big( d^{(\alpha-1)/\alpha} \Big) \Bigg]
\\
& \overset{\text{(a)}}{=}
\lim_{\alpha \to 1} \bigg( \frac{ \mathrm{d} }{ \mathrm{d} \alpha } \frac{ \alpha-1 }{ \alpha } \bigg)^{-1} \cdot \Bigg( \frac{ \mathrm{d} }{ \mathrm{d} \alpha } \ln \Bigg[ \sum_{d \in \mathcal{D}_{q}} \varepsilon_{d} \cdot \Big( d^{(\alpha-1)/\alpha} \Big) \Bigg] \Bigg)
\\
& =
\lim_{\alpha \to 1} \bigg( \frac{ 1 }{ \alpha^{2} } \bigg)^{-1} \cdot \Bigg( \Bigg[ \frac{ \mathrm{d} }{ \mathrm{d} \alpha } \sum_{d \in \mathcal{D}_{q}} \varepsilon_{d} \cdot \Big( d^{(\alpha-1)/\alpha} \Big) \Bigg] \cdot \Bigg[ \sum_{d \in \mathcal{D}_{q}} \varepsilon_{d} \cdot \Big( d^{(\alpha-1)/\alpha} \Big) \Bigg]^{-1} \Bigg)
\\
& =
\lim_{\alpha \to 1} \bigg( \frac{ 1 }{ \alpha^{2} } \bigg)^{-1} \cdot \Bigg( \Bigg[ \sum_{d \in \mathcal{D}_{q}} \varepsilon_{d} \cdot (\ln d) \cdot \bigg( \frac{ \mathrm{d} }{ \mathrm{d} \alpha } \frac{ \alpha - 1 }{ \alpha } \bigg) \cdot \Big( d^{(\alpha-1)/\alpha} \Big) \Bigg] \cdot \Bigg[ \sum_{d \in \mathcal{D}_{q}} \varepsilon_{d} \cdot \Big( d^{(\alpha-1)/\alpha} \Big) \Bigg]^{-1} \Bigg)
\\
& =
\lim_{\alpha \to 1} \bigg( \frac{ 1 }{ \alpha^{2} } \bigg)^{-1} \cdot \Bigg( \Bigg[ \sum_{d \in \mathcal{D}_{q}} \varepsilon_{d} \cdot (\ln d) \cdot \bigg( \frac{ 1 }{ \alpha^{2} } \bigg) \cdot \Big( d^{(\alpha-1)/\alpha} \Big) \Bigg] \cdot \Bigg[ \sum_{d \in \mathcal{D}_{q}} \varepsilon_{d} \cdot \Big( d^{(\alpha-1)/\alpha} \Big) \Bigg]^{-1} \Bigg)
\\
& =
\bigg( \frac{ 1 }{ 1^{2} } \bigg)^{-1} \cdot \Bigg( \Bigg[ \sum_{d \in \mathcal{D}_{q}} \varepsilon_{d} \cdot (\ln d) \cdot \bigg( \frac{ 1 }{ 1^{2} } \bigg) \cdot \Big( d^{(1-1)/1} \Big) \Bigg] \cdot \Bigg[ \sum_{d \in \mathcal{D}_{q}} \varepsilon_{d} \cdot \Big( d^{(1-1)/1} \Big) \Bigg]^{-1} \Bigg)
\\
& =
\Bigg[ \sum_{d \in \mathcal{D}_{q}} \varepsilon_{d} \cdot (\ln d) \Bigg] \cdot \Bigg[ \sum_{d \in \mathcal{D}_{q}} \varepsilon_{d} \Bigg]^{-1}
\\
& =
\sum_{d \in \mathcal{D}_{q}} \varepsilon_{d} \cdot (\ln d) ,
\label{eq:I1_V} \\
I_{\infty}(V)
& \overset{\eqref{def:alpha_infty}}{=}
\lim_{\alpha \to \infty} I_{\alpha}(V)
\\
& \overset{\eqref{eq:Ialpha_V}}{=}
\lim_{\alpha \to \infty} \frac{ \alpha }{ \alpha-1 } \ln \Bigg[ \sum_{d \in \mathcal{D}_{q}} \varepsilon_{d} \cdot \Big( d^{(\alpha-1)/\alpha} \Big) \Bigg]
\\
& =
\bigg( \underbrace{ \lim_{\alpha \to \infty} \frac{ \alpha }{ \alpha-1 } }_{ = 1 } \bigg) \cdot \Bigg( \lim_{\alpha \to \infty} \ln \Bigg[ \sum_{d \in \mathcal{D}_{q}} \varepsilon_{d} \cdot \Big( d^{(\alpha-1)/\alpha} \Big) \Bigg] \Bigg)
\\
& =
\ln \Bigg( \sum_{d \in \mathcal{D}_{q}} \varepsilon_{d} \cdot d \Bigg) ,
\label{eq:Iinfty_V}
\end{align}
where (a) follows by L'H{\^o}pital's rule.

We now consider the channel $V$ which the input alphabet size $q$ is a prime power.
If $\alpha \in (0, 1)$, then Jensen's inequality shows
\begin{align}
2 \, I_{\alpha}( V )
& \overset{\eqref{eq:Ialpha_V}}{=}
2 \times \left( \frac{ \alpha }{ \alpha-1 } \ln \Bigg[ \sum_{d \in \mathcal{D}_{q}} \varepsilon_{d} \cdot \Big( d^{(\alpha-1)/\alpha} \Big) \Bigg] \right)
\\
& \overset{\eqref{eq:conservation_eps}}{=}
2 \times \left( \frac{ \alpha }{ \alpha-1 } \ln \Bigg[ \sum_{d \in \mathcal{D}_{q}} \bigg( \frac{ \varepsilon_{d}^{-} }{2} +  \frac{ \varepsilon_{d}^{+} }{2} \bigg) \cdot \Big( d^{(\alpha-1)/\alpha} \Big) \Bigg] \right)
\\
& =
2 \times \left( \frac{ \alpha }{ \alpha-1 } \ln \Bigg[ \frac{1}{2} \Bigg( \sum_{d \in \mathcal{D}_{q}}  \varepsilon_{d}^{-} \cdot \Big( d^{(\alpha-1)/\alpha} \Big) \Bigg) + \frac{1}{2} \Bigg( \sum_{d \in \mathcal{D}_{q}} \varepsilon_{d}^{+} \cdot \Big( d^{(\alpha-1)/\alpha} \Big) \Bigg) \Bigg] \right)
\\
& \le
2 \times \left( \frac{ \alpha }{ \alpha-1 } \Bigg( \frac{1}{2} \ln \Bigg[ \sum_{d \in \mathcal{D}_{q}}  \varepsilon_{d}^{-} \cdot \Big( d^{(\alpha-1)/\alpha} \Big) \Bigg] \Bigg) + \Bigg( \frac{1}{2} \ln \Bigg[ \sum_{d \in \mathcal{D}_{q}} \varepsilon_{d}^{+} \cdot \Big( d^{(\alpha-1)/\alpha} \Big) \Bigg] \right)
\label{ineq:Jensen} \\
& =
\frac{ \alpha }{ \alpha-1 } \ln \Bigg[ \sum_{d \in \mathcal{D}_{q}} \varepsilon_{d}^{-} \cdot \Big( d^{(\alpha-1)/\alpha} \Big) \Bigg] + \frac{ \alpha }{ \alpha-1 } \ln \Bigg[ \sum_{d \in \mathcal{D}_{q}} \varepsilon_{d}^{+} \cdot \Big( d^{(\alpha-1)/\alpha} \Big) \Bigg]
\\
& \overset{\eqref{eq:Ialpha_V}}{=}
I_{\alpha}(V^{0}) + I_{\alpha}(V^{1}) ,
\end{align}
which is \eqref{eq:ineq_1} for $\alpha \in (0, 1)$.
Similarly, since the inequality of \eqref{ineq:Jensen} is reversed if $\alpha \in (1, \infty)$, the inequality \eqref{eq:ineq_2} also holds for $\alpha \in (1, \infty)$.

Finally, we consider relations among $I_{\alpha}(V)$, $I_{\alpha}(V^{0})$, and $I_{\alpha}(V^{1})$ for $\alpha \in \{ 0, 1, \infty \}$.
It follows from \eqref{eq:E-_primepower} and \eqref{eq:E+_primepower} that
\begin{align}
\sgn( \varepsilon_{d} )
=
\sgn( \varepsilon_{d}^{-} )
=
\sgn( \varepsilon_{d}^{+} )
\end{align}
for $d \in \mathcal{D}_{q}$, where $\sgn : \mathbb{R} \to \{ -1, 0, 1 \}$ denotes the sign function, i.e.,
\begin{align}
\sgn( x )
\coloneqq
\begin{cases}
1
& \mathrm{if} \ x > 0 , \\
0
& \mathrm{if} \ x = 0 , \\
-1
& \mathrm{if} \ x < 0 .
\end{cases}
\end{align}
Hence, it can be seen from \eqref{eq:I0_V} that
\begin{align}
I_{0}( V ) = I_{0}( V^{0} ) = I_{0}( V^{1} ) ,
\\
2 \, I_{0}( V ) = I_{0}( V^{0} ) + I_{1}( V^{1} ) .
\end{align}
Moreover, simple calculations yield
\begin{align}
2 \, I_{1}(V)
& \overset{\eqref{eq:I1_V}}{=}
2 \sum_{d \in \mathcal{D}_{q}} \varepsilon_{d} \cdot (\ln d)
\\
& \overset{\eqref{eq:conservation_eps}}{=}
2 \sum_{d \in \mathcal{D}_{q}} \bigg( \frac{\varepsilon_{d}^{-}  + \varepsilon_{d}^{+}}{2} \bigg) \cdot (\ln d)
\\
& =
\sum_{d \in \mathcal{D}_{q}} \varepsilon_{d}^{-} \cdot (\ln d) + \sum_{d \in \mathcal{D}_{q}} \varepsilon_{d}^{+} \cdot (\ln d)
\\
& \overset{\eqref{eq:I1_V}}{=}
I_{1}( V^{0} ) + I_{1}( V^{1} ) ,
\label{eq:conservation_ordinary_V} \\
2 \, I_{\infty}(V)
& \overset{\eqref{eq:Iinfty_V}}{=}
2 \ln \Bigg( \sum_{d \in \mathcal{D}_{q}} \varepsilon_{d} \cdot d \Bigg)
\\
& \overset{\eqref{eq:conservation_eps}}{=}
2 \ln \Bigg( \sum_{d \in \mathcal{D}_{q}} \bigg( \frac{ \varepsilon_{d}^{-} }{ 2 }  + \frac{ \varepsilon_{d}^{+} }{ 2 } \bigg) \cdot d \Bigg)
\\
& =
2 \ln \Bigg( \frac{1}{2} \Bigg( \sum_{d \in \mathcal{D}_{q}} \varepsilon_{d}^{-} \cdot d \Bigg) + \frac{1}{2} \Bigg( \sum_{d \in \mathcal{D}_{q}} \varepsilon_{d}^{+} \cdot d \Bigg) \Bigg)
\\
& \overset{\text{(a)}}{\ge}
2 \, \Bigg( \frac{1}{2} \ln \Bigg( \sum_{d \in \mathcal{D}_{q}} \varepsilon_{d}^{-} \cdot d \Bigg) + \frac{1}{2} \ln \Bigg( \sum_{d \in \mathcal{D}_{q}} \varepsilon_{d}^{-} \cdot d \Bigg) \Bigg)
\\
& =
\ln \Bigg( \sum_{d \in \mathcal{D}_{q}} \varepsilon_{d}^{-} \cdot d \Bigg) + \ln \Bigg( \sum_{d \in \mathcal{D}_{q}} \varepsilon_{d}^{+} \cdot d \Bigg)
\\
& \overset{\eqref{eq:Iinfty_V}}{=}
I_{\infty}( V^{0} ) + I_{\infty}( V^{1} ) ,
\end{align}
where (a) follows by Jensen's inequality.
Note that the identity \eqref{eq:conservation_ordinary_V} is well known as the conservation property of the (ordinary) symmetric capacity under the polar transformation.
This completes the proof of \corref{cor:ineq}.
\end{IEEEproof}

In \corref{cor:ineq}, it holds that $I( V^{0} ) + I( V^{1} ) = 2 I( V )$, which is well known as the conservation property of the the (ordinary) symmetric capacity (cf. \eqref{eq:conservation_ordinary_V}) under the polar transformation.
By the identity \eqref{eq:I_E0} and the change of variable as $\rho = (1 - \alpha ) / \alpha$, both of the inequalities \eqref{eq:ineq_1} and \eqref{eq:ineq_2} can be combined as
\begin{align}
E_{0}( \rho, V^{0} ) + E_{0}( \rho, V^{1} )
\ge
2 E_{0}( \rho, V )
\end{align}
for $\rho \in (-1, \infty)$, which is a similar result to \cite[Theorem~1]{alsan2}.

\begin{figure}[!t]
\centering
\begin{overpic}[width = 0.9\hsize, clip]{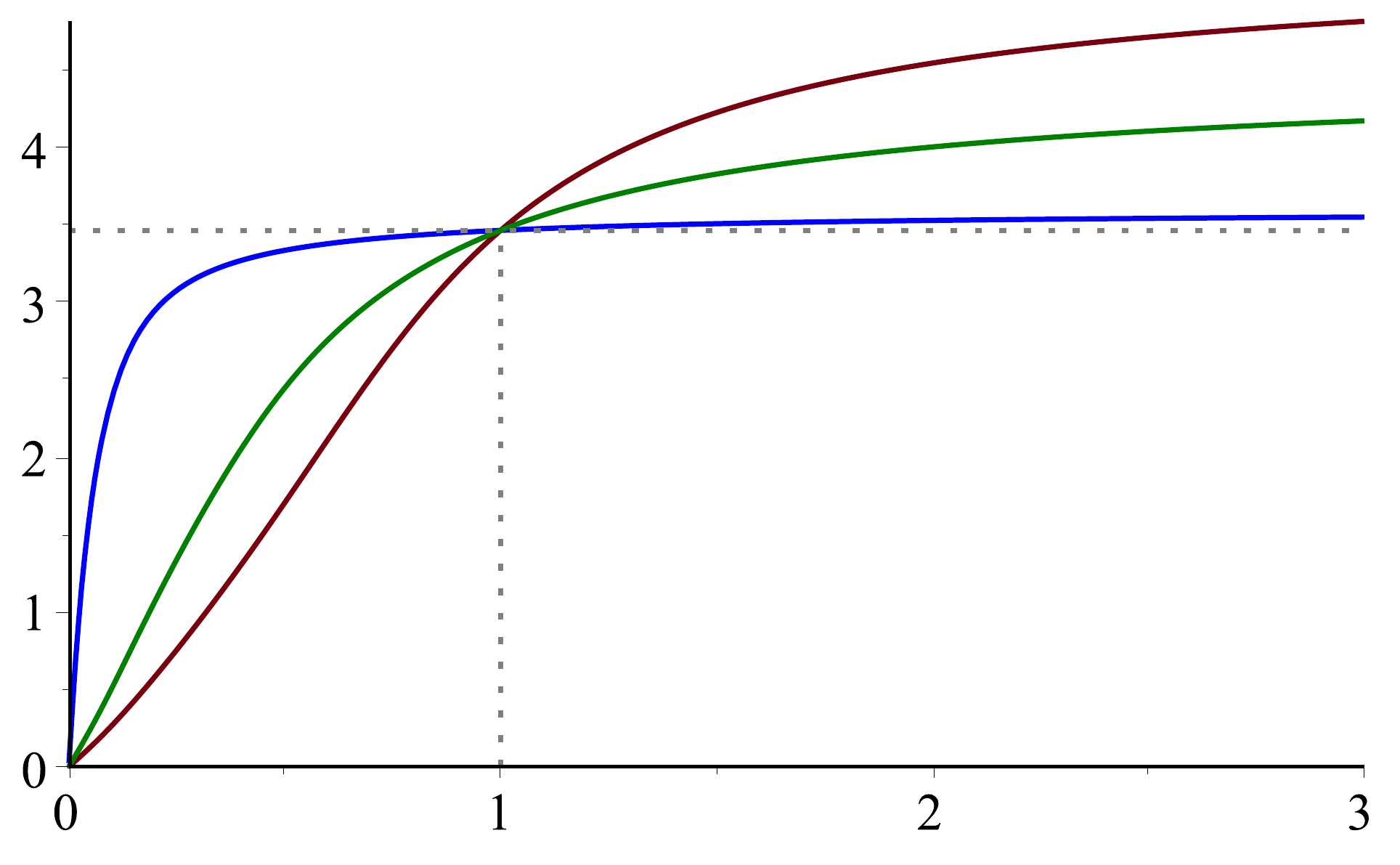}
\put(50, -0.5){order $\alpha$}
\put(-4, 30){\rotatebox{90}{$\mathbb{E}[ I_{\alpha}( V_{n} ) ]$}}
\put(-4, 58){\scriptsize [nats]}
\put(11, 52.5){$I_{1}(V) = 5 \ln 2$}
\put(10, 53){\vector(-2, -3){4.7}}
\put(50, 13.7){$\alpha = 1$ (symmetric capacity)}
\put(49, 13.7){\vector(-2, -1){12.5}}
\put(87, 36){\color{burgundy} $n = 0$, i.e., $I_{\alpha}(V)$}
\put(90, 39){\color{burgundy} \vector(-1, 4){5}}
\put(20, 10){\color{blue} $n = 8$}
\put(20, 13){\color{blue} \vector(-1, 2){11}}
\put(60, 28){\color{Green} $n = 2$}
\put(62, 31){\color{Green} \vector(1, 4){5}}
\end{overpic}
\caption{Average $\mathbb{E}[ I_{\alpha}( V_{n} ) ]$ of the symmetric capacity of order $\alpha$ for the random variable $V_{n}$, defined in \eqref{def:processV}.
The initial channel $V_{0} = V$ is given as follows: The input alphabet size is $q = 1024 = 2^{10}$, and the $11$-dimensional probability vector is $( \varepsilon_{d} : d \in \mathcal{D}_{1024} ) = (\varepsilon_{1}, \varepsilon_{2}, \varepsilon_{4}, \varepsilon_{8}, \varepsilon_{16}, \varepsilon_{32}, \varepsilon_{64}, \varepsilon_{128}, \varepsilon_{256}, \varepsilon_{512},  \varepsilon_{1024}) = (1/11, 1/11, \dots, 1/11)$, i.e., the uniform distribution.} 
\label{fig:alpha}
\end{figure}
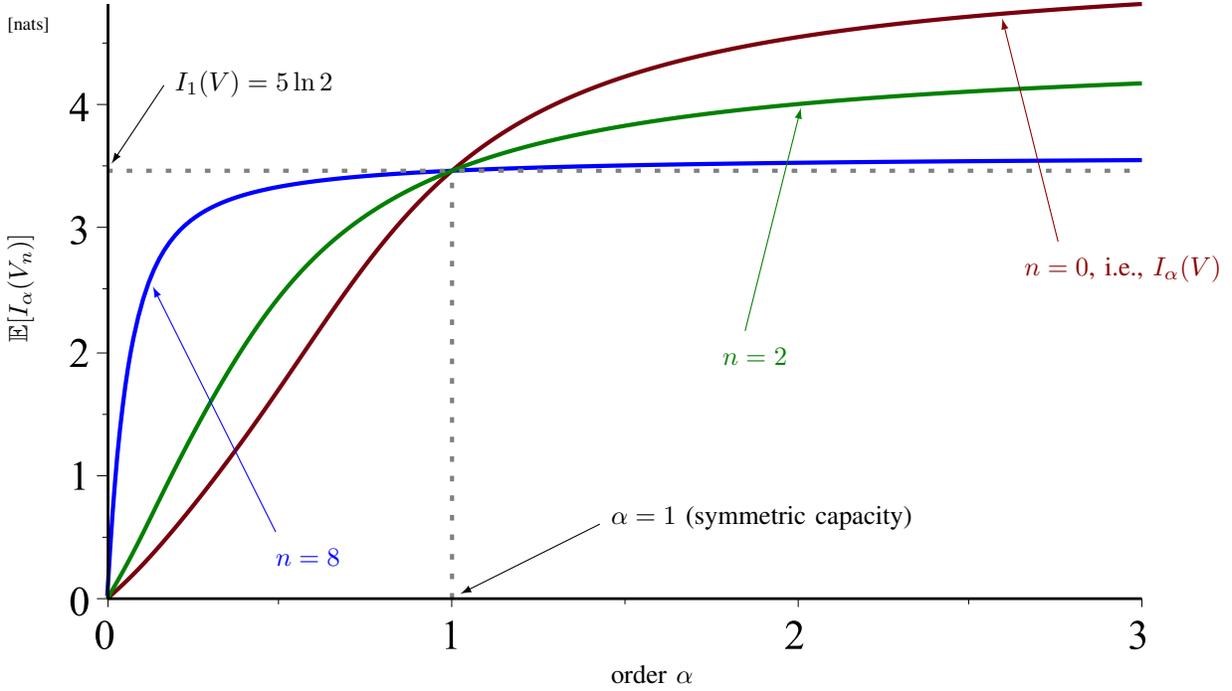

We now consider the polarization process of the channel $V$ as follows:
Define a Bernoulli process $(B_{n} : n \in \mathbb{N}) = (B_{1}, B_{2}, \dots)$, where $B_{n} \sim \mathrm{Bernoulli}( 1/2 )$ is a $\{ 0, 1 \}$-valued random variable for each $n \in \mathbb{N}$.
In the mapping \eqref{def:f}, suppose that $\gamma \in \mathbb{Z}_{q}^{\times}$.
Then, for an initial channel $V$, the polarization process $(V_{n} : n \in \mathbb{N}_{0})$ is defined by
\begin{align}
V_{n}
& \coloneqq
\begin{cases}
V
& \mathrm{if} \ n = 0 , \\
V_{n-1}^{B_{n}}
& \mathrm{if} \ n \ge 1 .
\end{cases}
\label{def:processV}
\end{align}
Namely, for each $n \in \mathbb{N}_{0}$, the random variable $V_{n}$ takes channels of \defref{def:V} uniformly from the set $\{ V_{2^{n}}^{(i)} \mid 0 \le i < 2^{n} \}$, which is generated by \eqref{def:Wi}.
Figure~\ref{fig:alpha} illustrates an average behavior of $I_{\alpha}( V_{n} )$ for $\alpha \in [0, 3]$, which is an graphical representation of \corref{cor:ineq}.
In \figref{fig:alpha}, note that $I_{\alpha}( W )$ is nondecreasing for $\alpha \in [0, \infty]$ (cf. \cite[part~4 of Lemma~1]{arimoto2}).
For the process $(V_{n} : n \in \mathbb{N}_{0})$, we now present the following theorem.

\begin{theorem}
\label{th:martingale}
Consider the polarization process $(V_{n} : n \in \mathbb{N})$ with an initial channel $V_{(q)}( \varepsilon_{d} : d \in \mathcal{D}_{q} )$.
If the input alphabet size $q$ is a prime power, then the random variable $V_{n}$ convergences almost surely to $V_{\infty}$ such that
\begin{align}
\Pr \Big( \forall \alpha \in [0, \infty] , \ I_{\alpha}( V_{\infty} ) = \ln d \Big) = \varepsilon_{d}
\end{align}
for $d \in \mathcal{D}_{q}$.
\end{theorem}

\begin{IEEEproof}[Proof of \thref{th:martingale}]
Suppose throughout the proof that $q = p^{m}$ for some prime $p$ and some $m \in \mathbb{N}$, i.e., the input alphabet size $p$ is a prime power.
Since $\mathcal{D}_{q} = \{ p^{i} \mid 0 \le i \le m \} = \{ 1, p, p^{2}, \dots, p^{m} \}$ when $p$ is a prime power, we see that the channel $V : \mathbb{Z}_{q} \to \mathcal{Y}$ is specified by an $(m+1)$-dimensional probability vector $(\varepsilon_{d} : d \in \mathcal{D}_{q} ) = (\varepsilon_{p^{r}} : 0 \le r \le m) = (\varepsilon_{1}, \varepsilon_{p}, \varepsilon_{p^{2}}, \dots, \varepsilon_{p^{m}})$.

Let $V_{(q)}(\varepsilon_{p^{r}} : 0 \le r \le m)$ be an initial channel.
We define an independent and identically distributed (i.i.d.) random variables $(S_{n} : n \in \mathbb{N}) = (S_{1}, S_{2}, \dots)$, where $S_{n}$ is a $\{ -, + \}$-valued uniformly distributed random variable for each $n \in \mathbb{N}$.
Employing the random variables $(S_{n} : n \in \mathbb{N})$, we now consider a stochastic process $(\bvec{E}_{n} : n \in \mathbb{N}_{0})$ of the polar transformation, calculated by \eqref{eq:E-_primepower} and \eqref{eq:E+_primepower}, as
\begin{align}
\bvec{E}_{n}
& \coloneqq
\begin{cases}
(\varepsilon_{p^{r}} : 0 \le r \le m)
& \mathrm{if} \ n = 0 , \\
\bvec{E}_{n-1}^{S_{n}}
& \mathrm{if} \ n \ge 1 ,
\end{cases}
\label{def:process_E}
\end{align}
where
$
\bvec{E}_{n}
=
(E_{r, n} : 0 \le r \le m)
=
(E_{0,n}, E_{1,n}, E_{2,n}, \dots, E_{m,n})
$
is an $(m+1)$-dimensional random vector and $\bvec{E}_{n-1}^{S_{n}} \coloneqq (E_{r, n-1}^{S_{n}} : 0 \le r \le m)$.
Namely, the random vector $\bvec{E}_{n}$ is recursively calculated by
\begin{align}
E_{r, n}
\coloneqq
E_{r, n-1}^{S_{n}}
=
\begin{cases}
\displaystyle
E_{r, n-1} \cdot \Bigg( E_{r, n-1} + 2 \sum_{t = r+1}^{m} E_{t, n-1} \Bigg)
& \mathrm{if} \ S_{n} = - , \\
\displaystyle
E_{r, n-1} \cdot \Bigg( E_{r, n-1} + 2 \sum_{t = 0}^{r-1} E_{t, n-1} \Bigg)
& \mathrm{if} \ S_{n} = +
\end{cases}
\label{eq:E_recursive}
\end{align}
for $r = 0, 1, 2, \dots, m$ and $n \in \mathbb{N}$.
Note that $V_{n} \equiv V_{(q)}( E_{r, n} : 0 \le r \le m )$ for $n \in \mathbb{N}_{0}$, where $V_{n}$ is defined in \eqref{def:processV}.
Since $P_{S_{n}}( - ) = P_{S_{n}}( + ) = 1/2$ for $n \in \mathbb{N}$, it follows that
\begin{align}
\mathbb{E}[E_{r, n} \mid S_{1}, S_{2}, \dots, S_{n-1}]
& =
\mathbb{E} \Big[ E_{r, n-1}^{S_{n}} \ \Big| \ S_{1}, S_{2}, \dots, S_{n-1} \Big]
\\
& =
\frac{1}{2} E_{r, n-1}^{-} + \frac{1}{2} E_{r, n-1}^{+}
\\
& \overset{\eqref{eq:conservation_eps}}{=}
E_{r, n-1}
\end{align}
for $r = 0, 1, 2, \dots, m$ and $n \in \mathbb{N}$, where $\mathbb{E}[ \cdot \mid \cdot ]$ denotes the conditional expectation of the random variable.
Thus, the stochastic process $(E_{r, n} : n \in \mathbb{N}_{0})$ is a martingale with respect to the sequence $(S_{n} : n \in \mathbb{N})$ for each $r = 0, 1, 2, \dots, m$.
It is also easy to see that $E_{r, n}$ is bounded in $\mathcal{L}^{s}$ for $0 \le s < \infty$, i.e., $\mathbb{E} [ |E_{r, n}|^{s} ] < \infty$ for $r = 0, 1, 2, \dots, m$ and $n \in \mathbb{N}$, since the random vector $\bvec{E}_{n}$ takes probability vectors for each $n \in \mathbb{N}$, i.e., $0 \le E_{r, n} \le 1$ almost surely for each $r = 0, 1, \dots, m$ and $n \in \mathbb{N}$.
Hence, the sequence $(E_{r, n} : n \in \mathbb{N}_{0})$ is uniformly integrable for each $r = 0, 1, \dots, m$; and therefore, the martingale $(E_{r, n} : n \in \mathbb{N}_{0})$ with respect to the sequence $(S_{n} : n \in \mathbb{N})$ convergences almost surely and in $\mathcal{L}^{1}$ to a random variable $E_{r, \infty}$ for each $r = 0, 1, 2, \dots, m$.
For this random vector $\bvec{E}_{\infty} = (E_{r, \infty} : 0 \le r \le m)$, the following lemma holds.

\begin{lemma}
\label{lem:martingale}
The random variable $E_{r, \infty}$ only takes either $0$ or $1$ almost surely for each $r = 0, 1, 2, \dots, m$.
\end{lemma}

\begin{IEEEproof}[Proof of \lemref{lem:martingale}]
For random variables $X$ and $Y$, let
\begin{align}
\mathbb{E}[ X ; Y = y ]
\coloneqq
\mathbb{E} \Big[ X \, \1[ Y = y ] \Big]
\end{align}
for $y \in \mathcal{Y}$, where $\mathcal{Y}$ is the range of $Y$, $\mathbb{E}[ \cdot ]$ denotes the expectation of the random variable, and
\begin{align}
\1 [ Y = y ]
\coloneqq
\begin{cases}
1
& \mathrm{if} \ Y = y , \\
0
& \mathrm{otherwise}
\end{cases}
\end{align}
is the indicator function.
If $E_{t, n} = 0$ for $t < r$ with a fixed $r = 0, 1, \dots, m$, then \eqref{eq:E_recursive} can be written as
\begin{align}
E_{r, n+1}
=
E_{r, n}^{S_{n+1}}
=
\begin{cases}
2 E_{r, n} - E_{r, n}^{2}
& \mathrm{if} \ S_{n+1} = - , \\
E_{r, n}^{2}
& \mathrm{if} \ S_{n+1} = +
\end{cases}
\label{eq:E_recursive_2}
\end{align}
since $\sum_{t=r}^{m} E_{t, n} = 1$ when $E_{t, n} = 0$ for $t = 0, 1, \dots, r-1$.
Then, we get
\begin{align}
&
\mathbb{E} \Big[ \, \Big| E_{r, n+1} - E_{r, n} \Big| \, ; \: \forall t < r , \ E_{t, n} = 0 \Big]
\notag \\
& \qquad =
\mathbb{E} \Big[ \, \Big| E_{r, n}^{S_{n+1}} - E_{r, n} \Big| \, ; \: \forall t < r , \ E_{t, n} = 0 \Big]
\\
& \qquad =
\frac{1}{2} \mathbb{E} \Big[ \, \Big| E_{r, n}^{-} - E_{r, n} \Big| \, ; \: \forall t < r , \ E_{t, n} = 0 \Big] + \frac{1}{2} \mathbb{E} \Big[ \, \Big| E_{r, n}^{+} - E_{r, n} \Big| \, ; \: \forall t < r , \ E_{t, n} = 0 \Big]
\\
& \qquad \overset{\eqref{eq:E_recursive_2}}{=}
\frac{1}{2} \mathbb{E} \Big[ \, \Big| 2 E_{r, n} - E_{r, n}^{2} - E_{r, n} \Big| \, ; \: \forall t < r , \ E_{t, n} = 0 \Big] + \frac{1}{2} \mathbb{E} \Big[ \, \Big| E_{r, n}^{2} - E_{r, n} \Big| \, ; \: \forall t < r , \ E_{t, n} = 0 \Big]
\\
& \qquad =
\frac{1}{2} \mathbb{E} \Big[ \, \Big| E_{r, n} - E_{r, n}^{2} \Big| \, ; \: \forall t < r , \ E_{t, n} = 0 \Big] + \frac{1}{2} \mathbb{E} \Big[ \, \Big| E_{r, n}^{2} - E_{r, n} \Big| \, ; \: \forall t < r , \ E_{t, n} = 0 \Big]
\\
& \qquad \overset{\text{(a)}}{=}
\frac{1}{2} \mathbb{E} \Big[ E_{r, n} - E_{r, n}^{2} \, ; \: \forall t < r , \ E_{t, n} = 0 \Big] + \frac{1}{2} \mathbb{E} \Big[ E_{r, n} - E_{r, n}^{2} \, ; \: \forall t < r , \ E_{t, n} = 0 \Big]
\\
& \qquad =
\mathbb{E} \Big[ E_{r, n} \big( 1 - E_{r, n} \big) ; \: \forall t < r , \ E_{t, n} = 0 \Big]
\\
& \qquad \overset{\text{(b)}}{\to}
0 \quad (\mathrm{as} \ n \to \infty) ,
\end{align}
where (a) follows from the fact that
$
0 \le E_{r, n} \le 1
$,
and (b) follows by the $\mathcal{L}^{1}$ convergence.
Thus, we observe that
\begin{align}
\mathbb{E} \Big[ E_{r, \infty} \big( 1 - E_{r, \infty} \big) ; \: \forall t < r, \ E_{t, \infty} = 0 \Big]
=
0
\end{align}
for each $r = 0, 1, \dots, m$, which implies that
\begin{align}
\Pr \Big( E_{r, \infty} = 0 \ \mathrm{or} \ E_{r, \infty} = 1 \ \Big| \ \forall t < r , \ E_{t, \infty} = 0 \Big)
& =
1 \quad \mathrm{for} \ r = 0, 1, \dots, m ,
\label{prob:Einfty_1}
\end{align}
where $\Pr( \cdot \mid \cdot )$ denotes the conditional probability.
Moreover, since the random variable $\bvec{E}_{\infty}$ takes probability vectors, i.e., $\sum_{r = 0}^{m} E_{r, \infty} = 1$ and $E_{t, \infty} \ge 0$ almost surely for all $t = 0, 1, \dots, m$, we have
\begin{align}
\Pr \Big( E_{r, \infty} = 0 \ \Big| \ \exists r^{\prime} \neq r \ \mathrm{s.t.} \ E_{r^{\prime}, \infty} = 1 \Big)
& =
1 \quad \mathrm{for} \ r = 0, 1, \dots, m .
\label{prob:Einfty_2}
\end{align}
Combining \eqref{prob:Einfty_1} and \eqref{prob:Einfty_2}, we have \lemref{lem:martingale}.
\end{IEEEproof}

Since $0 \le E_{r, \infty} \le 1$, it follows from \lemref{lem:martingale} that
\begin{align}
\Pr( E_{r, \infty} = 1 )
=
1 - \Pr( E_{r, \infty} = 0 )
=
\mathbb{E}[ E_{r, \infty} ]
\end{align}
for $r = 0, 1, \dots, m$.
Moreover, the property of the expectation of the martingale shows
\begin{align}
\mathbb{E}[ E_{r, \infty} ]
=
\mathbb{E}[ E_{r, 0} ]
\overset{\eqref{def:process_E}}{=}
\varepsilon_{p^{r}}
\end{align}
for $r = 0, 1, \dots, m$.
Hence, we have
\begin{align}
\Pr( E_{r, \infty} = 1 )
=
\varepsilon_{p^{r}}
\label{eq:Pr_E_r_infty}
\end{align}
for $r = 0, 1, \dots, m$, where (a) follows by \lemref{lem:martingale}.
Therefore, the random vector $\bvec{E}_{\infty}$ takes deterministic probability vectors.

From \eqref{eq:Ialpha_V}, \eqref{eq:I0_V}, \eqref{eq:I1_V}, and \eqref{eq:Iinfty_V}, since the channel $V \equiv V_{(q)}( \varepsilon_{d} : d \in \mathcal{D}_{q} )$ satisfies
$
I_{\alpha}( V ) = \ln d
$
for $\alpha \in [0, \infty]$ when $\varepsilon_{d} = 1$ for some $d \in \mathcal{D}_{q}$, we have
\begin{align}
\Pr \Big( \forall \alpha \in [0, \infty] , \ I( V_{\infty} ) = \ln \big( p^{r} \big) \Big)
& =
\Pr \Big( \forall \alpha \in [0, \infty] , \ I \big( V_{(q)}( E_{r, \infty} : 0 \le r \le m ) \big) = \ln \big( p^{r} \big) \Big)
\\
& \overset{\eqref{eq:Pr_E_r_infty}}{=}
\varepsilon_{p^{r}}
\end{align}
for $r = 0, 1, \dots, m$, which completes the proof of \thref{th:martingale}.
\end{IEEEproof}

After some algebra, it follows that 
\begin{align}
I_{\alpha}( V )
& =
\frac{ \alpha }{ \alpha - 1 } \ln \sum_{d \in \mathcal{D}_{q}} \varepsilon_{d} \Big( d^{(\alpha - 1)/\alpha} \Big)
\end{align}
for $\alpha \in (0, 1) \cup (1, \infty)$, and
\begin{align}
I_{0}( V )
& =
\min_{d \in \mathcal{D}_{q} : \varepsilon_{d} > 0} \Big( \ln d \Big) ,
\\
I_{1}( V )
& =
\sum_{d \in \mathcal{D}_{q}} \varepsilon_{d} \ln d ,
\label{eq:I(V)} \\
I_{\infty}( V )
& =
\ln \Bigg( \sum_{d \in \mathcal{D}_{q}} \varepsilon_{d} \cdot d \Bigg)
\end{align}
(cf. the proof of \corref{cor:ineq}).
Therefore, we obtain
\begin{align}
\forall \alpha \in [0, \infty], \ I_{\alpha}( V ) = \ln d
\iff
\varepsilon_{d} = 1
\end{align}
for $d \in \mathcal{D}_{q}$, which implies that the random variable $V_{\infty}$ takes partially noiseless channels.
We now check that, if $\varepsilon_{d} = 1$ for some $d \in \mathcal{D}_{q}$, then the channel $V$ is partially noiseless as follows:
For a given $q$-ary input channel $V : \mathbb{Z}_{q} \to \mathcal{Y}$, we define the $d$-ary input degenerated channel $V[d] : \mathbb{Z}_{d} \to \mathcal{Y}$ as
\begin{align}
V[d](y \mid x)
\coloneqq
\frac{ d }{ q } \sum_{x^{\prime} \in \mathbb{Z}_{q} \cap [x]_{d}} V(y \mid x^{\prime})
\end{align}
for $d \in \mathcal{D}_{q}$.
If $\varepsilon_{d} = 1$, then we readily see that
$
I( V[d] )
=
\ln d
$
and
$
P_{\mathrm{e}}( V[d] )
=
0
$,
which implies a $d$-ary input noiseless channel.
In addition, it can be seen that the zero-error capacity \cite{zero} of the channel $V$ is $\ln d$ when $\varepsilon_{d} = 1$, where the consideration is related to the study by Guo et al. \cite{guo}.

Furthermore, \thref{th:martingale} ensures the proportion of the multilevel polarization of $I_{\alpha}( V_{\infty} )$.
Roughly speaking, the proportion $|\{ 0 \le i < 2^{n} \mid I_{\alpha} \big( V_{2^{n}}^{(i)} \big) \approx \ln d \}| / 2^{n}$ is nearly equal to $\varepsilon_{d}$ for $d \in \mathcal{D}_{q}$ when $n$ is sufficiently large (cf. \figref{fig:V27}).

\section{Conclusion}

In this study, we proposed a new class of arbitrary input generalized erasure channels $V$ in \defref{def:V}, which contains BECs, OECs \cite[p.~2285]{park1}, and \cite[Fig.~4: Channel~2]{sahebi}.
For the proposed channel $V$, \thref{th:V} established recursive formulas of the polar transformation $V \overset{\gamma}{\mapsto} (V^{0}, V^{1})$.
In \sectref{sect:prime_power}, we analyzed the polarization of $I_{\alpha}( V )$ in more detail when the input alphabet size $q$ is a prime power, and stated \thref{th:martingale}.
\thref{th:martingale} ensured the proportion of the multilevel polarization of $\{ I_{\alpha}( V_{2^{n}}^{(i)} ) \mid 0 \le i < 2^{n} \}$ with $n$ sufficiently large (cf. \figref{fig:V27}).

Finally, we remark that \thref{th:V} and Corollaries~\ref{cor:V} and~\ref{cor:ineq} can be easily extended to the polar transformation with two independent (but not necessarily identical) channels $V$ and $V^{\prime}$, which is defined in, e.g., \cite[Eqs.~(9) and~(10)]{alsan2}.







\begin{thebibliography}{99}


\bibitem{alsan2}
M.~Alsan and E.~Telatar,
``Polarization improves $E_{0}$,''
\emph{IEEE\ Trans.\ Inf.\ Theory},
vol.~60, no.~5, pp.~2714--2719, May 2014.

\bibitem{arikan}
E.~Ar{\i}kan,
``Channel polarization: A method for constructing capacity-achieving codes for symmetric binary-input memoryless channels,''
\emph{IEEE\ Trans.\ Inf.\ Theory},
vol.~55, no.~7, pp.~3051--3073, July 2009.

\bibitem{arimoto}
S.~Arimoto,
``On the converse to the coding theorem for discrete memoryless channels,''
\emph{IEEE\ Trans.\ Inf.\ Theory},
vol.~19, no.~3, pp.~357--459, May 1973.

\bibitem{arimoto2}
---------,
``Information measures and capacity of order $\alpha$ for discrete memoryless channels,''
in \emph{Topics in Information Theory, 2nd Colloq. Math. Soc. J. Bolyai},
Keszthely, Hungary, vol.~16, pp.~41--52, 1977.




\bibitem{red}
R.~G.~Gallager,
\emph{Information Theory and Reliable Communication.}
New~York: Wiley, 1968.

\bibitem{guo}
J.~Guo, J.~Sayir, M.~Qin, and A.~Guill{\'e}n~i~F{\`a}bregas,
``An alternative proof of channel polarization for channels with arbitrary input alphabets,''
\emph{Proc.\ 53rd\ Annual\ Allerton\ Conf.\ Commun.,\ Control,\ Comput.},
Monticello, IL, USA, Sept.--Oct. 2015.


\bibitem{ho}
S.-W.~Ho and S.~Verd{\'u},
``Convexity/concavity of R{\'e}nyi entropy and $\alpha$-mutual information,''
\emph{Proc.\ IEEE\ Int.\ Symp.\ Inf.\ Theory} (ISIT'2015),
Hong~Kong, pp.~745--749, June 2015.




\bibitem{mackay}
D.~J.~C.~MacKay,
\emph{Information Theory, Inference, and Learning Algorithms.}
Cambridge: Cambridge University Press, 2003.

\bibitem{mori}
R.~Mori and T.~Tanaka,
``Source and channel polarization over finite fields and Reed-Solomon matrices,''
\emph{IEEE\ Trans.\ Inf.\ Theory},
vol.~60, no.~5, pp.~2720--2736, May 2014.

\bibitem{nasser1}
R.~Nasser,
``Ergodic theory meets polarization I: A foundation of polarization theory,''
\emph{Proc.\ IEEE\ Int.\ Symp.\ Inf.\ Theory} (ISIT'2015),
Hong~Kong, pp.~2451--2455, June 2015.

\bibitem{nasser2}
---------,
``Ergodic theory meets polarization II: A foundation of polarization theory for MACs,''
\emph{Proc.\ IEEE\ Int.\ Symp.\ Inf.\ Theory} (ISIT'2015),
Hong~Kong, pp.~2456--2460, June 2015.

\bibitem{nasser3}
R.~Nasser and E.~Telatar,
``Polarization theorems for arbitrary DMCs,''
\emph{Proc.\ IEEE\ Int.\ Symp.\ Inf.\ Theory} (ISIT'2013),
Istanbul, Turkey, pp.~1297--1301, July 2013.

\bibitem{park1}
W.~Park and A.~Barg,
``The ordered Hamming metric and ordered symmetric channels,''
\emph{Proc.\ IEEE\ Int.\ Symp.\ Inf.\ Theory} (ISIT'2011),
St.~Peterburg, Russia, pp.~2283--2287, Aug. 2011.

\bibitem{park2}
---------,
``Polar codes for $q$-ary channels, $q = 2^{r}$,''
\emph{IEEE\ Trans.\ Inf.\ Theory},
vol.~59, no.~2, pp.~955--969, Feb. 2013.

\bibitem{sahebi_arxiv}
A.~G.~Sahebi and S.~S.~Pradhan,
``Multilevel polarization of polar codes over arbitrary discrete memoryless channels,''
July 2011. [Online].
Available at \url{http://arxiv.org/abs/1107.1535}.

\bibitem{sahebi}
---------,
``Multilevel channel polarization for arbitrary discrete memoryless channels,''
\emph{IEEE\ Trans.\ Inf.\ Theory},
vol.~59, no.~12, pp.~7839--7857, Dec. 2013.

\bibitem{sasoglu2}
E.~{\c{S}}a{\c{s}}o{\u{g}}lu,
``Polar codes for discrete alphabets,''
\emph{Proc.\ IEEE\ Int.\ Symp.\ Inf.\ Theory} (ISIT'2012),
Cambridge, MA, USA, pp.~2137--2141, July 2012.

\bibitem{sasoglu}
E.~{\c{S}}a{\c{s}}o{\u{g}}lu, E.~Telatar, and E.~Ar{\i}kan,
``Polarization for arbitrary discrete memoryless channels,''
\emph{Proc.\ IEEE\ Inf.\ Theory\ Workshop} (ITW'2009),
Taormina, Sicily, Italy, pp.~144--148, Oct. 2009.

\bibitem{shannon}
C.~E.~Shannon,
``A mathematical theory of communication,''
\emph{Bell Syst. Tech. J.},
vol.~27, pp.~379--423~and~623--656, July~and~Oct. 1948.

\bibitem{zero}
---------,
``The zero error capacity of a noisy channel,''
\emph{IRE\ Trans.\ Inf.\ Theory},
vol.~2, no.~3, pp.~8--19, Sept. 1956.

\bibitem{sphere}
C.~E.~Shannon, R.~G.~Gallager, and E.~R.~Berlekamp,
``Lower bounds to error probability for coding on discrete memoryless channels. I,''
\emph{Inf.\ Control},
vol.~10, no.~1, pp.~65--103, Jan. 1967.

\bibitem{tal}
I.~Tal and A.~Vardy,
``How to construct polar codes,''
\emph{IEEE\ Trans.\ Inf.\ Theory},
vol.~59, no.~10, pp.~6562--6582, Oct. 2013.

\bibitem{verdu}
S.~Verd\'{u},
``$\alpha$-mutual information,''
\emph{Proc.\ IEEE\ Inf.\ Theory\ Appl.\ Workshop} (ITA'2015),
CA, USA, pp.~1--6, Feb. 2015.

\end{thebibliography}
%

\end{document}